\documentstyle[epsf,rotate,psfig]{mn_plus_bib}

\def\HST{{\it HST}}

\def\Mpc{\,\hbox{Mpc}}

\def\Msun{\hbox{M}_{\odot}}

\def\kms{\,\hbox{km}\,\hbox{s}^{-1}}

\def\kpc{\,\hbox{kpc}}
\def\H0{H_0=100 \, h \, {\rm kms^{-1}Mpc^{-1}}}

\def\SN{S_N}

\def\Gyr{\hbox{Gyr}}

\def\Galform{\textsc{galform}}
\def\kms{kms$^{-1}$}

\def\etal{{et al.\thinspace}}
\def\eg{{e.g.\thinspace}}

\def\ie{{i.e.\thinspace}}

\def\U{{\emph{U}}}
\def\B{{\emph{B}}}
\def\V{{\emph{V}}}

\def\K{{\emph{K}}}

\def\gsim{~\rlap{$>$}{\lower 1.0ex\hbox{$\sim$}}}

\begin{document}
\title{On the Formation of Globular Cluster Systems in a Hierarchical Universe}

\author[M.A.~Beasley et al.]
{M.A.~Beasley,$^{1}$\thanks{email:
mbeasley@astro.swin.edu.au} C.M.~Baugh,$^2$ Duncan
A.~Forbes,$^1$ R.M.~Sharples,$^2$ C.S.~Frenk$^2$\\ 
  $^1$Centre for Astrophysics \& Supercomputing, 
  Swinburne University of Technology, Hawthorn, VIC 3122, Australia.\\
  $^2$Department of Physics, University of Durham, Durham DH1 3LE, UK.}

\date{Accepted~~~~~~~~~~.   Received~~~~~~~~~~.}

\pagerange{\pageref{firstpage}--\pageref{lastpage}}
\pubyear{2002}

\label{firstpage}

\maketitle

\begin{abstract}
We have investigated the formation of globular 
cluster (GC) systems in the fiducial semi-analytic 
model of galaxy formation of Cole et al., by assuming 
that GCs are formed at high-redshift ($z >$ 5)
in proto-galactic fragments, and during the 
subsequent gas-rich merging of these fragments. 
Under these assumptions  
we have simulated the GC systems of 450 
elliptical galaxies, and find 
that the majority (93 \%) are intrinsically 
bimodal in metallicity. 
We find that, in the mean, the metal-rich 
GC sub-populations are younger than the metal-poor 
GC sub-populations, with ages of 9 Gyr and 12 Gyr respectively, 
and that the mean ages of the metal-rich GCs are 
dependent upon host galaxy luminosity and 
environment (halo circular velocity), whereas the metal-poor 
GCs are not. 
We find that the continued gaseous merging
of the proto-galactic fragments leads to significant 
age-structure amongst the metal-rich GCs.
These GCs exhibit a large age-range (5 to 12 Gyr), 
which increases for low-luminosity galaxies, and 
for galaxies in low circular velocity haloes.
Moreover, the metal-rich GCs associated with low-luminosity field and/or 
group ellipticals, are $\sim$ 2 Gyr younger than the metal-rich
GCs in luminous cluster ellipticals.
We find the total GC populations scale with host galaxy 
luminosity as $N_{\rm GC} \propto L_{V, \rm
gal}^{1.25}$, a result in agreement with observations 
of luminous elliptical galaxies.
This scaling is due to a systematic increase 
in the $\mathcal{M}/L$ ratios of the galaxy haloes with
luminosity for L$_{V, \rm gal} > $ L$_*$ galaxies in the model.
A comparison between the luminosity growth of the model ellipticals
and their GC formation indicates that mergers do not
significantly effect $\SN$ at $z < 2$.
We find the mean colours of the both the metal-rich and metal-poor
GCs exhibit only a weak dependence upon host galaxy luminosity, 
a result consistent with contemporary observations.
We conclude that gaseous merging, the bulk of which occurs at
1 $\leq z \leq$ 4 in our $\Lambda$CDM model, leads to the
formation of the metal-rich peak of the GC systems of elliptical
galaxies. We suggest that the formation and subsequent
truncation of the metal-poor GCs in the proto-galactic fragments 
is closely related to the star formation rate in these fragments,
which may have been significantly higher at very early times.
\end{abstract}


\section{Introduction}
\label{sec:Introduction}

Globular clusters (GCs) are a seemingly ubiquitous 
feature of nearly all galaxies in the local 
Universe.
A growing body of evidence 
has indicated the presence of correlations
between GC systems and their host galaxies, 
such as the mean metallicity of GC systems
(\eg \citeANP{Brodie91} 1991)
and the total number of GCs with host
galaxy luminosity (\eg \citeANP{Djorgovski92} 1992;
\citeANP{Zepf94} 1994) 
or cluster velocity dispersion 
(\eg \citeANP{Blakeslee97} 1997).
Results such as these suggest that GC formation 
is intimately linked to galaxy star formation, 
and since GCs are simpler stellar systems than galaxies, 
they provide a unique record of the formation history of their
host galaxy.

In the past decade, there have been 
two key advances in GC research.
One has been the discovery of newly
forming, massive young star clusters
(YMCs to use the terminology of 
\citeANP{Larsen99} 1999), with masses and sizes
consistent with those expected for newly formed GCs.
Such YMCs are seen to be forming in
a variety of star-bursting environments \cite{Meurer95}
including violent galaxy mergers (\eg 
\citeANP{Holtzman92} 1992; \citeANP{Whitmore95} 1995;
\citeANP{Schweizer96} 1996; \citeANP{Zepf99} 1999;
\citeANP{Forbes00a} 2000), 
star-bursts (\eg \citeANP{OConnell94} 1994;
\citeANP{Watson96} 1996), circum-nuclear ring galaxies
\cite{Barth95} and the discs of otherwise unperturbed
spirals \cite{Larsen99}.
The inference is that YMCs can form in a 
wide variety of galactic environments.

The second development has been the discovery
of bimodality in the broad-band colours of many
(perhaps most) GC systems (\eg \citeANP{Zepf93} 1993;
\citeANP{Elson96} 1996; \citeANP{Geisler96} 1996; 
\citeANP{Forbes97a} 1997;   
\citeANP{Gebhardt99} 1999; \citeANP{Kundu01} 2001; 
\citeANP{Larsen01} 2001).
Owing to the relative insensitivity
of broad-band colours to age differences in old
stellar populations (\citeANP{OConnell76} 1976;
\citeANP{Worthey94} 1994) these bimodal colour
distributions in GC systems are, by implication, 
thought to largely reflect bimodal metallicity 
distributions. Spectroscopic metallicities
obtained for the Milky Way 
(see \citeANP{Harris01} 2001), M31 \cite{Barmby00} and several
giant ellipticals (\eg \citeANP{Cohen98} 1998;
\citeANP{KisslerPatig98} 1998;  
\citeANP{Beasley00} 2000; \citeANP{Forbes01b} 2001) 
largely support this notion.

The metallicity bimodality of the GC systems
in luminous ellipticals was predicted by \citeANP{Ashman92} (1992), 
who presented a model for the formation of metal-rich
GCs during the merging of gas-rich spirals.
Subsequent ideas have also been developed which 
attempt to explain the bimodality (and other properties)
of GC systems,  {\it in-situ}, 
multi-phase collapse (\citeANP{Forbes97} 1997;   
\citeANP{Harris98} 1998) and the accretion of
dwarf galaxies and their GCs \cite{Cote98}.
Whilst all of these models have been 
successful in explaining certain properties of 
GC systems, in detail each has its problems 
(see \citeANP{Harris01} 2001 for details).  
However, rather than (somewhat 
artificially) separating these
models, it is perhaps more
reasonable to expect a hybrid of some
or all of the above scenarios. 
In this respect, 'semi-analytic' models of galaxy 
formation may offer a suitable framework within 
which to achieve such a hybrid model in a more natural way.

Semi-analytic models of galaxy formation 
have proved themselves to be valuable 
tools with which to follow the formation and 
subsequent evolution of galaxies in hierarchical
clustering cosmologies 
(\eg \citeANP{White78} 1978; \citeANP{WhiteFrenk91};
\citeANP{Kauffmann93} 1993; \citeANP{Cole94} 1994). 
These models treat the galaxy formation problem in 
an {\it ab initio} fashion.
Starting from a power spectrum of primordial density fluctuations, 
the merger history and assembly of dark matter haloes is computed.
The properties of the galaxy population are then calculated 
through to the present day using simple, physically motivated 
analytic recipes to follow the evolution of gas and stars.

The parameters and input physics of the semi-analytic models 
are constrained by the requirement that the models should match
various `fundamental' galaxy properties, and these constraints
vary between the different groups. Such properties 
include the local \B\ and \K-band galaxy luminosity functions, 
the relative fractions of galaxy morphological types, 
the slope of the Tully-Fisher relation and 
the gas fractions in discs in the case of the model used in 
this paper (\eg \citeANP{Cole94} 1994; 2000; \citeANP{Baugh96} 1996; 1998), 
or the zero-point of the Tully-Fisher relation and the slope of the 
galaxy colour-magnitude relation in the case of the Munich and Santa Cruz 
groups (\eg \citeANP{Kauffmann93} 1993; \citeANP{Kauffmann98} 1998; 
\citeANP{Somerville99} 1999).
The resulting model predictions are then compared to 
other observed galaxy properties.

Currently, none of the semi-analytic 
models have been tested in any 
detail against the growing wealth of data on GC systems of galaxies. 
Here, for the first time, we compare expectations from the 
fiducial semi-analytic model of \citeANP{Cole00} (2000) with
the observed properties of GC systems. 

\begin{figure}
\epsfysize 3.4truein
\hfil{\epsffile{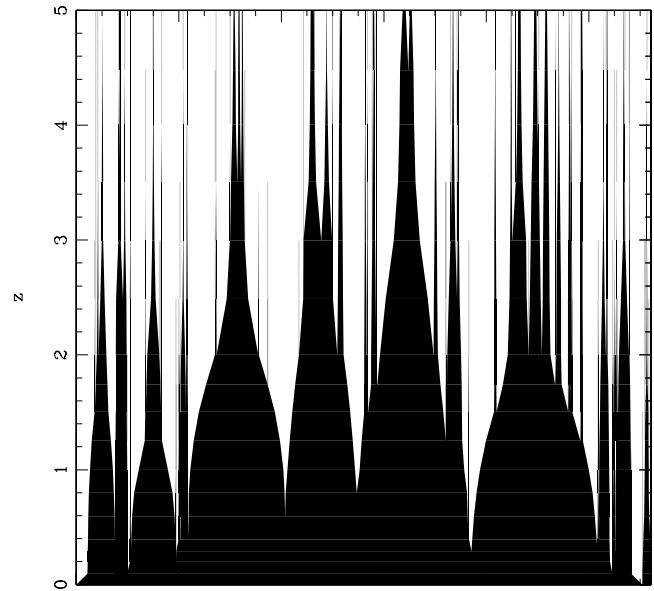}}\hfil
\caption{Merger tree for an elliptical galaxy of mass
$\sim 2.3 \times 10^{11}~\Msun$, beginning at redshift 5, 
and proceeding up to the present day. 
The width of the 'branches' reflect mass at a
given epoch. The merger tree has been normalised to possess
unit width at $z$ = 0.}
\label{fig:tree}
\end{figure}

The layout of this paper is as follows: 
in Section~\ref{sec:Methodology}
we describe the semi-analytic model used in this study
and how we go about producing GC systems from
the star formation histories it produces.
In Section~\ref{sec:TheGlobularClusterSystemsofIndividualGalaxies}
we show examples of the resulting GC systems of 
individual galaxies and compare them with 
observations of GC systems.
Next, in Section~\ref{sec:GlobalPropertiesoftheGlobularClustersystems},
we compare the global properties 
of our model GC systems with a compilation of contemporary 
observational data for the GC systems of early-type galaxies.
In Section~\ref{sec:Discussion} we discuss some
of the principal issues raised in this study, before
presenting a summary and our conclusions from this
work in Section~\ref{sec:SummaryandConclusions}.

\section{Methodology}
\label{sec:Methodology}

\subsection{The Semi-Analytic Model}
\label{subsec:GALFORM}

The semi-analytic model we use in this study
is the \Galform code (\citeANP{Cole00} 2000).
A wide range of physical processes are incorporated 
in this model; here we give only a brief outline of the 
model and refer the reader to \citeANP{Cole00} (2000) 
for a complete description.

The \Galform\ model used in this paper is the 
fiducial $\Lambda$CDM model of \citeANP{Cole00}(2000), with 
the following values of the cosmological parameters: 
a mean mass density, $\Omega_0$ = 0.3, a cosmological
constant, $\Lambda_0$ = 0.7, a  Hubble constant 
of $h=0.7$ (where $H_0$ = 100 $h$ \kms \Mpc$^{-1}$) 
and a linear variance in spheres of radius 
$8h^{-1}$Mpc of $\sigma_{8}=0.93$.  
Once the cosmological parameters and the form of the 
power spectrum of density fluctuations have been 
specified, the full merger history of an ensemble of 
dark matter haloes is computed using the 
Monte-Carlo prescription described in \citeANP{Cole00}(2000) 
(see also \citeANP{Baugh98} 1998).

The semi-analytic code utilises a set of simple rules to 
to model the complex physics of galaxy formation. The processes 
treated include: {\it (i)} The shock heating of gas in the gravitational 
potential wells of dark matter haloes. {\it (ii)} The subsequent 
radiative cooling of this gas, tracking the  
angular momentum of the gas distribution. {\it (iii)} The formation 
of stars from the cooled gas. {\it (iv)} Feedback processes, e.g. 
stellar winds and supernovae, that re-heat some of the cooled gas 
and thereby regulate the star formation rate. 
{\it (v)} The chemical evolution of the various reservoirs of 
gas and stars (see Fig. 3 of \citeANP{Cole00} 2000). 
{\it (vi)} Mergers of galaxies inside a common dark matter halo, 
based upon dynamical friction arguments. Bursts of star formation 
can occur during a merger if cold gas is present.
{\it (vii)} The evolution of the stellar populations of galaxies. 
The extinction of starlight is computed in a self-consistent 
way using the chemical evolution model and the size of the 
disk and bulge components. 
We note that Benson \etal (2001a) describe extensions to the scheme of 
\citeANP{Cole00}(2000), refining the treatment of the cooling 
of gas in relatively small dark matter haloes at high redshift 
and improving the calculation of merger timescales. These 
features are not used in the model presented here. For bright 
galaxies at reasonable redshifts, we expect that these two versions 
of the \Galform\ code will produce comparable predictions.

We show an example of the evolutionary 
history of a \Galform\ galaxy in Figure~\ref{fig:tree} 
(see \citeANP{Baugh96} 1996, \citeANP{Baugh98} 1998). 
The figure shows the formation and evolution of the 
progenitors of a present day massive elliptical galaxy. 
The width of the filled regions indicates the fraction 
of the present day galaxy's stellar mass that is contributed  
by each progenitor.
In the case of Figure~\ref{fig:tree}, the progenitors of this 
elliptical galaxy experienced several fairly significant mergers 
that would lead to a dramatic change in the progenitors' morphology 
(\eg at redshifts 0.8 and 0.3), and numerous minor mergers.

\begin{figure}
\epsfysize 3.4truein
\hfil{\epsffile{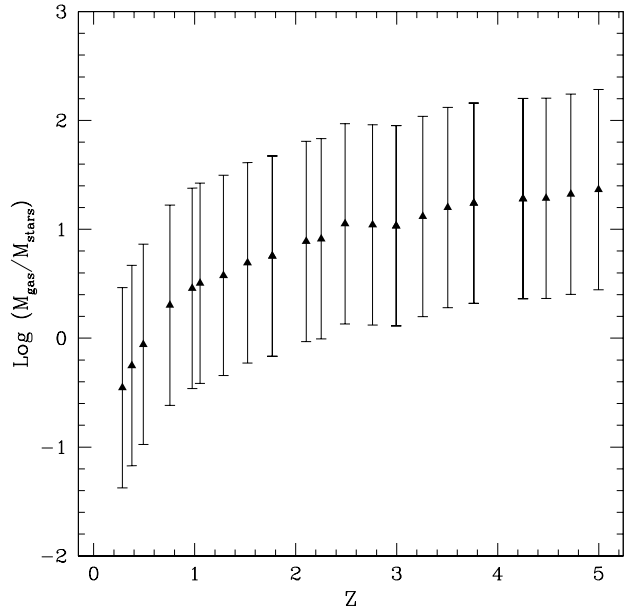}}\hfil
\caption{Mean ratio of the mass in cold gas to stars for the 
progenitor PGDs for a present day elliptical 
in the semi-analytic model. Filled triangles
with error-bars represent mean values with their
associated 10 and 90 percentiles of the distribution. 
The cold gas-fractions with respect to stars range 
from $\sim$ 95\% at $z$ = 5, to $\sim$ 10\% at $z$ = 0.}
\label{fig:coldgas}
\end{figure}

Star formation in the discs of the progenitors of present day 
galaxies provides the source of metal-poor GCs in our 
model (see next section).
When we say GCs are formed in gas discs, 
this does {\it not} mean
that the metal-poor GCs are associated with the stellar/gas discs
of spiral galaxies. Rather that the origin of the
metal-poor GCs are proto-galactic cold-gas 
discs modelled in \Galform. Henceforth, to distinguish between 
galaxy discs and the proto-galactic discs 
modelled in \Galform, we abbreviate the latter 
to {\it PGDs.}

We show the evolution of the mean ratio of cold gas 
to stars in the PGDs of a present-day 
elliptical in Figure~\ref{fig:coldgas}.
At redshift $\sim$ 5, the majority of these PGDs
have a high mass fraction of cold gas ($\sim$ 95\%).
It is only at $ z \sim$ 0.6 that this ratio reaches unity, 
as gas is used up during star formation.

In addition to star formation in PGDs, bursts of 
star formation can accompany galaxy mergers. 
Galaxies merge in the model if the timescale for a 
galaxy orbit to decay to the centre of a dark matter 
halo, as a result of dynamical friction, is shorter 
than the lifetime of the halo, which we take to be the 
time for the halo to double its mass through the  
accretion and mergers of smaller subunits.
If the mass of the satellite is larger than 
$30\%$ of the mass of the central galaxy, then 
the merger is designated as a major merger 
(see \citeANP{Kauffmann93} 1993; \citeANP{Baugh96}1996).
In this case, if cold gas is present then a burst 
of star formation takes place.
Thus it can be seen that the morphology of a galaxy in 
\Galform\ is essentially a function of time.
In a major merger, any stellar discs in the merging galaxies
are destroyed and the resulting galaxy will
be bulge dominated, \ie it will be early-type.
Over time, in the absence of additional major mergers, gas will 
cool and the bulge will accrete
a disc of enriched, cold gas, thus forming 
a spiral (or lenticular) galaxy.
An important point to bear in mind when 
considering this model is that much of the star formation 
occurs at relatively low redshift, 
with only $\sim$ 50 \% of stars forming
prior to redshifts of $z \sim$ 1.5. This occurs
because the formation of stars in lower mass haloes is
suppressed by feedback from star formation. 

\subsection{Star Formation Histories to Globular Clusters}
\label{StarFormationHistoriestoGlobularClusters}

\subsubsection{Forming Globular Clusters}
\label{subsubsec:FormingGlobularClusters}

Our goal is to create a simple model for 
forming GC systems within \Galform, 
thereby keeping the number of new
parameters required to a minimum.
However, in creating GC systems from the 
star formation histories 
(SFHs) produced in the model, requires
us to make several assumptions about the nature 
of GC formation and how this relates
to the global star formation in a galaxy.
Our principal assumption is that GC
formation accompanies the two modes of star
formation. 
We assume that GCs form: 
{\it (i)} contemporaneously with the star formation
occurring in PGDs and 
{\it (ii)} during major mergers involving star formation.

The GCs formed in both modes
become the composite GC system of the
resulting galaxy. As we show later, this
can result in two sub-populations of GCs
which have undergone different levels of
chemical enrichment; one metal-poor
GC sub-population, formed in cold gas PGDs,
and one metal-rich formed during gas-rich merging.
Henceforth, in order to distinguish between the 
GC sub-populations, we refer to the 
PGD-formed GCs as $blue$ GCs and
the merger-formed GCs as $red$ GCs.

For each model realisation,
we build the GC system of each galaxy 
by assuming that some fraction of the stars
formed remain in a clustered mode, \ie\ they
become, and remain, GCs. 
We calculate total numbers of GCs (since we do not
have explicit luminosity information), and adopt 
a mean GC mass, $\langle M_{\rm GC}\rangle$, corresponding to that of the
Milky Way GC system of $\sim 3.0 \times 10^5 \Msun$
(from the McMaster catalogue of \citeANP{Harris96}
1996; May 15$^{\rm th}$, 1997 version).

We neglect the possible effects of
the dynamical evolution and destruction of the GCs,
which may be significant
(\eg \citeANP{Aguilar88}
1988; \citeANP{Gnedin97} 1997).
The study of \citeANP{McLaughlin99} (1999) suggests that 
the total mass of GC systems is not significantly effected
by the dynamical destruction of low-mass GCs
within an effective radius of an elliptical.
However, the simulations of \citeANP{Vesperini00} (2000) indicate
that some 10--20\%  of the initial GC system of a luminous
elliptical may be destroyed over a Hubble time due to dynamical
destruction processes.

Subsequently, we define a 'formation efficiency' of the GCs as
the mass of GCs which form and survive the possible effects of 
dynamical destruction ($M_{\rm GC}$) relative to the 
mass of stars which form and survive ($M_{\rm stars}$) 
associated with the model galaxies: 

\begin{equation}
\label{eq:efficiency}
\epsilon \equiv \frac{M_{\rm GC}}{M_{\rm stars}}
\end{equation}

\noindent it is effectively this efficiency -- the fraction of 
GCs that are formed and survive with respect to stars -- which is
measured when one obtains the 'specific frequency' ($\SN$) of a galaxy, 
as discussed further in $\S$~\ref{subsec:SpecificFrequency}.

Under the hypothesis that {\it no} dynamical destruction of the
GCs operates, this efficiency may be directly equated to a global
formation efficiency for the GCs. If, however, GC systems truly
undergo significant dynamical destruction
(e.g. \citeANP{Vesperini00} 2000), then $\epsilon$ is more a
reflection of a GC 'survival efficiency'.

With our stated assumptions, we construct
GC systems for each galaxy as follows:
for each galaxy merger tree, 
\Galform\ outputs SFHs for stars
formed both in PGDs and in bursts, 
yielding stellar masses, 
hot gas mass, cold gas mass and 
metallicities (Z) 
as a function of look-back time. 
To create the blue GCs, we 
take a fraction of the cumulative total of stars
formed in the PGDs, dictated by $M_{\rm GC}$ 
and our adopted blue GC formation efficiency ($\epsilon_{\rm blue}$).
This yields age and metallicity distributions 
for the metal-poor GCs as a function
of redshift. 

For each major merger,  
red GCs are created corresponding to our adopted
red GC formation efficiency ($\epsilon_{\rm red}$), providing 
that a high enough mass of stars are formed in the burst
($M_{*}$ $> \langle M_{\rm GC}\rangle$). 
These nascent GCs are then assigned 
ages and metallicities corresponding to 
the age and metallicity of the burst in which they
were formed.
Each burst of star formation is assumed 
to be instantaneous, and each burst assumes
instantaneous recycling of stellar material.
Therefore, the metallicity distributions of 
the red GCs form a series of 
$\delta$-functions, separated by the time elapsed
between each burst.\footnote{Our assumption 
of instantaneous GC formation is not strictly correct. Observational 
evidence (\eg \citeANP{Whitmore99} 1999)
and theoretical work \cite{Mihos96}
on such induced star formation indicates that 
its duration may be of order 
$\sim$ 100 Myr.}

We then convert the ages and metallicities of each GC
into observable quantities.
The absolute metallicity, $Z$, predicted by \Galform\
is converted to the amount of metals relative
to hydrogen via [m/H] = log$(Z/Z_{\odot}$), 
assuming $Z_{\odot}$ = 0.019 \cite{Anders89}.
In the observational plane, metallicity is often
expressed as the ratio of iron-peak elements 
relative to hydrogen, [Fe/H], and 
Galactic halo stars and GCs are known to be 
$\alpha$-enhanced with respect to solar values
by $\sim$ + 0.3 \cite{Harris01}.
Therefore, where appropriate, we convert between [m/H] and [Fe/H] 
using the approximate relation [Fe/H] = [m/H] -- 0.3.

Broad-band colours, such as $V-I$ and $V-K$, are then derived
for the GCs using the predicted ages and metallicities
of our model.
For this purpose, we use the single stellar population (SSP) models of 
\citeANP{Kurth99} (1999; henceforth KFF99).
We have investigated the effects of using the KFF99 models 
which adopt both Salpeter and Scalo initial mass
functions (IMFs). We find altering the assumed IMF has 
little effect on the {\it overall} colour distributions of the GC
systems, and for the remainder of this study we use the KFF99
models which employ a Salpeter IMF.

\subsubsection{Selecting the Initial Parameters}
\label{subsubsec:SelectingtheInitialParameters}

The principal parameters which we adjust 
are the formation efficiencies of the red and 
blue GCs. 
Several workers have argued for a 
'universal' efficiency of GC formation
(\eg \citeANP{diFazio87} 1987; \citeANP{McLaughlin96} 1996;
\citeANP{McLaughlin99} 1999), 
whilst other studies indicate that the star and GC
formation efficiency in star-bursts and interactions may be significantly
higher than in the 'passive' mode 
(\eg  \citeANP{Meurer95} 1995; \citeANP{Zepf99} 1999; 
\citeANP{Larsen00} 2000; \citeANP{Ashman01} 2001).
Since the exact physical processes governing 
GC formation are still unknown (\eg \citeANP{Elmegreen97} 1997), 
rather than adopting {\it a priori} assumptions
in any one particular direction, we have elected to employ
a straightforward parametrisation for the GC formation.

We have chosen to adjust the GC formation
efficiencies in our model in order to
reproduce the numbers of GCs observed in the prototypical
bimodal GC system of NGC~4472, 
perhaps the best-studied elliptical galaxy in
terms of its GC system.
This galaxy, an E2/S0 in the Virgo cluster with
$M_B$ = --22.05, has some 4000 blue GCs and 2400 
red GCs (\citeANP{Geisler96} 1996; \citeANP{Lee98} 1998), 
yielding a ratio of blue to red clusters
of $N_{\rm blue} / N_{\rm red} = 1.6$.
We have selected an elliptical in our simulated sample
(see $\S$~\ref{subsubsec:TheGalaxySample})
which offers the closest match to NGC~4472
in terms of luminosity, morphology
and halo mass.

The galaxy in question, \#12 in our model sample,  
is a giant elliptical with $M_B$ -- 5 log $h$ = --22.04, 
and a dark matter halo rotational velocity ($V_{\rm rot}$) 
of 600 \kms.
Assuming an isothermal potential and
isotropic velocity dispersion, we convert $V_{\rm rot}$
to 1-D velocity dispersions, $\sigma_{\rm 1D} = v_{\rm c} / \sqrt{2}$
\cite{Baugh96}. This yields $\sigma_{\rm 1D} =$ 424 \kms,
slightly higher, but comparable to that of the NGC~4472
subcluster in Virgo, where $\sigma_{\rm 1D} =$ 390 \kms
\cite{Binggeli87}. Henceforth, we will refer to this 
model elliptical as our fiducial galaxy.

In \Galform, the (globally averaged) luminous star formation rate (SFR) 
reaches a maximum of $\sim$ 0.1 $h$ $\Msun$ yr$^{-1}$ Mpc$^{-3}$
at redshift $\sim$ 3, and is still approximately 15\% of 
this value at the present epoch.
If we adopt a common efficiency for both GC components, 
and allow blue GC formation to continue up to $z$ = 0
in un-merged PGDs, the blue GCs dominate (by mass)
over the red GCs by up to two orders of magnitude
(depending upon the merger history of the
parent galaxy).

\begin{figure}
\epsfysize 3.4truein
\hfil{\epsffile{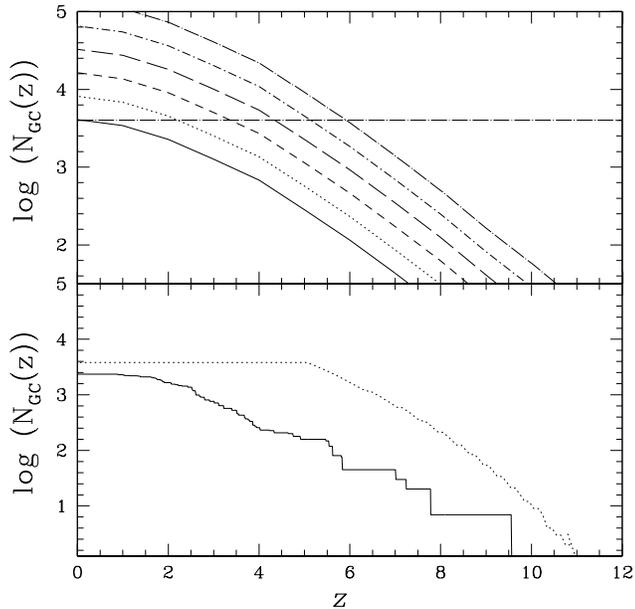}}\hfil
\caption{Cumulative number of GCs formed with redshift
for our fiducial galaxy.
{\it Top panel:} logarithm of the number of 
blue GCs formed in our fiducial galaxy with redshift.
The different curves (left to right) indicate  
GC formation efficiencies of 0.00012 to 
0.0040.
The horizontal dot-dashed line indicates the observed number 
of blue GCs in NGC~4472. {\it Bottom panel:} comparison
of the number of blue GCs formed with redshift in PGDs (dotted line)
and red GCs in bursts (solid line), normalised
to the observed numbers of red and blue GCs 
in NGC~4472 at $z$ = 0. The efficiencies are $\epsilon_{\rm blue}$ = 0.002
and $\epsilon_{\rm red}$ = 0.007, and the blue GC formation
has been truncated at $z$ = 5. }
\label{fig:truncate}
\end{figure}

To illustrate this effect,
in Figure~\ref{fig:truncate} (top panel)
we plot the number of blue GCs formed 
with redshift for our fiducial galaxy, 
for a range of GC formation efficiencies 
of 0.00012 to 0.004.
For comparison, we also show in the figure  
the number of GCs in the blue mode of
the NGC~4472 GC system ($\sim$ 4000), taken 
from \citeANP{Geisler96} (1996) (dot-dashed line).
It is clear that, unless we assume
a very low efficiency for GC formation, the 
PGD GC formation must be truncated at 
successively higher redshifts to 
obtain the correct total number of blue GCs.
In the bottom panel of Figure~\ref{fig:truncate}, we
show the cumulative total of blue GCs 
and red GCs formed with redshift.
We have adopted efficiencies of 
$\epsilon_{\rm blue}$ = 0.002 and 
$\epsilon_{\rm red}$ = 0.007
to agree with the observed
numbers of blue and red GCs in NGC~4472. 
The flattening of the curve for the blue GCs 
in Figure~\ref{fig:truncate} indicates the location of our 
truncation redshift.

\begin{figure}
\epsfysize 3.7truein
\hfil{\epsffile{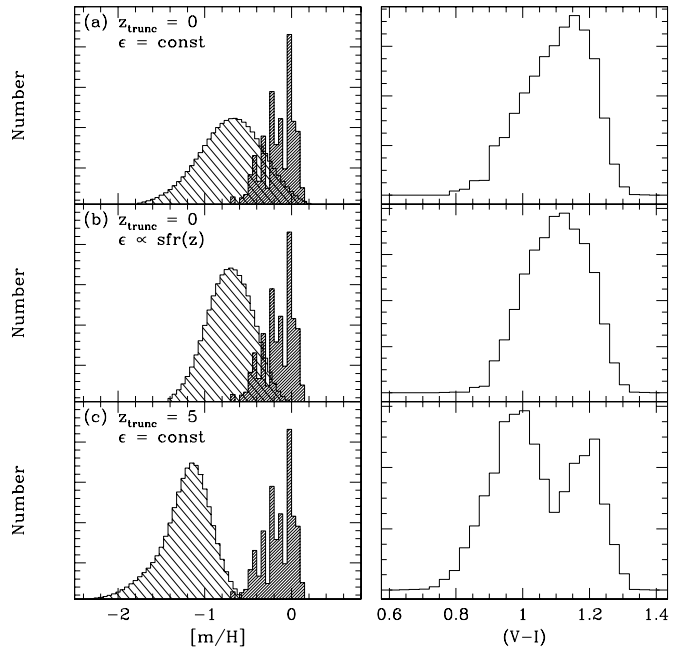}}\hfil
\caption{The resulting metallicity distributions (left panels)
and corresponding $V-I$ colour predictions (right panels)
for the red and blue GCs in our fiducial galaxy
under different assumptions. 
The distribution in
[m/H] of red and blue GCs if {\it (a)}
blue GC formation is allowed to continue to $z$ = 0, with
the efficiency of the blue GC formation assumed to be constant
at $\epsilon_{\rm blue}$ = 0.00012.
{\it (b)} blue GC formation continues to the present
epoch, but with $\epsilon_{\rm blue} \propto$ SFR and 
{\it (c)} 
$\epsilon_{\rm blue}$ = 0.002, but blue GC formation is
truncated at $z$ = 5.
The ages and metallicities are converted to colours using
the models of KFF99.}
\label{fig:feh}
\end{figure}

Upon initial consideration, it seems undesirable
to arbitrarily truncate the formation of the
blue GCs; we could simply choose a very low 
efficiency (\eg $\epsilon_{\rm blue}$ = 0.00012).
However, by allowing blue GC formation 
to continue to the 
present epoch significantly broadens the blue GC
metallicity distribution, to the extent that the 
observed colour distributions do not appear
bimodal. Such a result is inconsistent 
with the present observational situation, not
only for NGC~4472, but also for nearly all 
early-type galaxies with high-quality datasets. 

In Figure~\ref{fig:feh} we show  
the differential metallicity distributions (left-hand panels) 
for the red and blue GCs formed 
in our fiducial galaxy.
In producing the metallicity distributions of the
red GCs (dark shaded histograms in left-hand 
panels), we assume a constant formation efficiency of 
$\epsilon_{\rm red}$ = 0.007.
For the blue GCs we make three different assumptions 
about the nature of their formation efficiencies and 
formation redshift, each time ensuring that $\sim$ 4000
GCs are created. In the right-hand panels, 
we show the corresponding  $V-I$ colour distributions
of the GCs, predicted by interpolating the KFF99 models for
their corresponding metallicities and ages.

In Figure~\ref{fig:feh}{\it (a)}, we assume that
blue GCs form in PGDs up until the present epoch, and 
adopt an efficiency of $\epsilon_{\rm blue}$ = 0.00012, 
yielding $\sim$ 4000 GCs in the blue mode.
The blue metallicity distribution covers the 
range --1.8 $\leq $ [m/H] $\leq$ 0.2, 
peaking at [m/H] $\sim$ --0.7, and
overlaps substantially with the red GCs, which peak 
just short of solar metallicity. The peak-to-peak separation is
$\sim$ 0.5 dex and it is this, coupled with the fact
that the distribution is so broad ($\sim$ 0.5 dex),
which yields a colour distribution which is single-peaked
and skewed to $V-I \sim$ 1.18.

In {\it (b)} we again allow blue GC formation to 
continue to the present day, however this time
we set $\epsilon_{\rm blue}$ to be
proportional to the SFR in their progenitor PGDs. 
We have based this assumption on the results
of \citeANP{Larsen00} (2000), who have performed 
observations of YMCs in 21 nearby spiral galaxies. 
These authors found that
the specific \U-band luminosity of YMCs is correlated 
with the host galaxy far-infrared luminosity.
This, they propose, is indicative of the formation efficiency 
of the clusters being dependant upon the local SFR 
in the disc, as opposed to any dependence upon the $type$ of
star formation (\eg merger, quiescent, star-burst).
By adopting $\epsilon \propto$ SFR, we find that the 
blue GC metallicity distribution becomes more sharply peaked
than is the case for {\it (a)} and is narrower, 
with $\sigma_{\rm rms}$ = 0.28 dex. However,
the mean of this distribution still occurs at 
[m/H] $\sim$ --0.7.
Likewise, the corresponding $V-I$ colours of the GCs
also show a narrower distribution ($\sigma_{\rm rms}$ = 0.22 mag), 
but remain unimodal. At the peak of the SFR in the 
semi-analytic model ($z \sim$ 3), mean
stellar metallicities are already substantially
enriched, therefore, in the present scheme, an 
$\epsilon \propto$ SFR relation yields blue GCs
which cover too broad a metallicity range, and are 
too metal-rich (if no blue GC truncation
is adopted).

In {\it (c)} we show the GC system formed in the
same galaxy if we adopt a constant GC formation efficiency
($\epsilon_{\rm blue}$ = 0.002), but truncate
the blue GC formation at $z \sim$ 5.
This redshift corresponds to a characteristic
mean gas surface-density of the progenitor disks
of $\sim$ 1000$\Msun$ pc$^{-2}$ 
(1.25 $\times 10^{23}$ cm$^{-2}$).
Such a truncation results in a blue GC distribution in the 
range --2.4 $\leq $ [m/H] $\leq$ --0.6, peaking
at [m/H] $\sim$ --1.2.
Since the metallicity of the red GC component
peaks at [m/H] $\sim$ --0.2, this yields a 
peak-to-peak separation of 1.0 dex, and a 
corresponding colour distribution 
which is bimodal, with peaks at $V-I$ = 0.95 and 1.18. 

\begin{table}
\begin{center}
\caption{Model parameters adopted in this study.}
\label{tab:parameters}
\begin{tabular}{lll}
\hline 
 Adopted Cosmology \\
\hline
$\Omega_0$ & 0.3\\
$\Omega_b$ & 0.02\\
$\Lambda_0$ & 0.7\\
$h$ & 0.7\\
$\Gamma$ & 0.19\\
$\sigma_{8}$ & 0.93\\
\hline
Mergers \\
\hline
$f_{\rm ellip}$ & 0.3\\
\hline
Star and GC Formation \\
\hline
IMF & Salpeter\\
$R$ & 0.336\\
$y$ & 0.02\\
$\langle M_{\rm GC}\rangle$ & 3.0 $\times 10^5~\Msun$\\
$\epsilon_{\rm blue}$ & 0.002 \\
$\epsilon_{\rm red}$ & 0.007\\
$z_{\rm trunc}$ & 5\\
\hline
\end{tabular}
\end{center}
\end{table}

From the constraints placed upon us by observations 
of the GC system of NGC~4472, we obtain
$\epsilon_{\rm blue}$ = 0.002 and 
$\epsilon_{\rm red}$ = 0.007 respectively.
Furthermore, we truncate the formation of the
blue GCs at $z$ = 5 (henceforth denoted $z_{\rm trunc}$), 
which corresponds to look-back times of 
$\sim$ 12.3 \Gyr\ in our assumed cosmology.
We find that $\epsilon_{\rm blue}$ is 
similar to the global value found by 
\citeANP{McLaughlin99} (1999)
of $\epsilon$ = 0.0026 $\pm$ 0.0005, whilst
$\epsilon_{\rm red}$, the GC
formation efficiency in mergers, is required
to be a factor of $\sim$ 4 higher to produce the
correct number of GCs.

We list the principal 
cosmological and model parameters adopted in 
this study in Table~\ref{tab:parameters}.
In Table~\ref{tab:parameters}, $f_{\rm ellip}$
refers to the fraction of satellite to central
galaxy mass required to induce a major merger (here, 30\%), 
$y$ is the yield assumed in calculating the 
metal-enrichment during star formation in PGDs and
$R$ is the fraction of mass recycled by stars.
Again we refer the reader to \citeANP{Cole00} (2000)
for a full description of the model, 
and the effects of varying one or more of these
parameters.

\subsubsection{The Galaxy Sample}
\label{subsubsec:TheGalaxySample}

We have simulated a total of 450 
dark matter haloes, with masses
1.0$\times 10^{13} h^{-1} \Msun$ --
1.3$\times 10^{15} h^{-1} \Msun$.
The associated 450 central galaxies 
(and their respective GC systems) span a magnitude
range of --23.6 $\leq M_V$ - 5 log $h \leq$ --20.0.
The most luminous galaxies are rather
rare constructs in \Galform\ (as they are in the Universe), 
and therefore the lowest luminosity galaxies are 
numerically by far the most dominant in our simulated
sample.

Our galaxies have bulge-to-total (B/T)
luminosity ratios of 0.6 -- 1.0. These B/T values 
roughly correspond to early-type galaxy 
morphologies (see \citeANP{Kauffmann93} 1993; 
\citeANP{Baugh96} 1996).
For convenience, 
in this paper we refer to 
low-luminosity ellipticals as those
galaxies with $M_V$ -- 5 log $h \geq$ --21.5, 
and define the division between field/group and cluster
environments to lie (somewhat arbitrarily) 
at $\sigma_{\rm 1D}$ = 300 \kms\ (\ie velocity dispersions similar
to that of the Leo group). \newline

We emphasise the term 'environment' as used here refers to the
velocity dispersion (mass) of the parent dark matter halo, rather
than any number density measurement of nearby
galaxies. Consequently, our definitions of field/group/cluster
environments are not directly comparable to the density ($\rho$)
parameter used by some previous workers.

\section{The Globular Cluster Systems of Individual Galaxies}
\label{sec:TheGlobularClusterSystemsofIndividualGalaxies}

\subsection{A Luminous Elliptical}
\label{subsec:ALuminousElliptical}

\begin{figure}
\epsfysize 3.4truein
\hfil{\epsffile{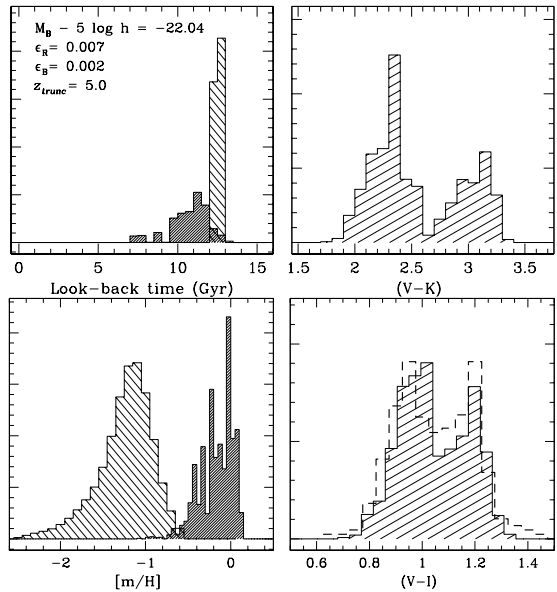}}\hfil
\caption{Age, metallicity and colour 
distributions for the GC system of our fiducial 
luminous elliptical galaxy.
{\it Top left panel:} 
Age (look-back time) distributions for blue GCs (light shaded histogram)
and red GCs (dark shaded histogram). 
{\it Bottom left panel:} Metallicity 
distributions for blue and red GCs.
{\it Top right:} Predicted $V-K$ colours of GCs converted 
using the stellar population models of KFF99.
{\it Bottom right:} $V-I$ colours of GCs using 
the KFF99 models.
The uncertainty assigned to the colours
is 0.05 mag. The dashed line indicates the scaled \HST\
data for NGC~4472 from Larsen \etal (2001).
}
\label{fig:galform.1}
\end{figure}

In Figure~\ref{fig:galform.1} we show the age, 
metallicity and corresponding colour ($V-I$ and $V-K$)
distributions obtained for our fiducial galaxy, 
the cluster elliptical with $M_B$ -- 5 log $h$ = --22.04.
Inspection of the age histograms (top left,
Figure~\ref{fig:galform.1}) 
indicates that the mean age of the blue GCs is several 
\Gyr\ older than the red GCs (although there
is some small overlap), and possess
a significantly shorter period of formation.
This short duration of blue GC formation is 
a result of our truncation at $z$ = 5, 
and the subsequent non-linear conversion 
between look-back time and redshift.
The more extended and continuous formation of 
the red GCs is a consequence of ongoing
merging and star formation 
from high redshift to $z \sim$ 0.6.

The bottom left panel of  
Figure~\ref{fig:galform.1} shows
the metallicity distribution
of these same GCs. The blue GCs 
show an exponential-like
increase in metallicity peaking
at [m/H] $\sim$ --1.2 and then a
sharp cut-off which, again, is a 
result of our truncation. 
The metallicity
range of these blue GCs is 
--2.7 $\leq$ [m/H] $\leq$ --0.5.
In contrast, the red GCs form
in multiple bursts at, or around, solar
metallicities (there are 
75 mergers involving star formation for
this particular galaxy).
Their metallicities
extend from [m/H] $\sim$ --1.0 to 
[m/H] $\sim$ + 0.2, with a 
mean metallicity of [m/H] = --0.2.
There is very little 
overlap between the metallicity
distributions of the two components.

The width of the blue and red
metallicity distributions are 0.33
and 0.20 dex respectively.
The model red GC distribution is somewhat 
narrower than those found 
from ground-based imaging by \citeANP{Geisler96} (1996) for NGC~4472, 
which are 0.35 and 0.36 dex for the blue and 
red modes respectively.  
However, these values may be overestimated
if the conversion between colour and metallicity
is non-linear (\eg\ see \citeANP{KisslerPatig98} 1998).
Our values are closer to those found for
the metal-poor and metal-rich components
of the Milky Way of 0.34 and 0.23 dex 
respectively \cite{AshmanandZepf98}.

In the top right panel
we show the 'observed' $V-K$ colours
of our fiducial galaxy's GC system, 
assuming observational uncertainties of 0.05 mag.
The distribution is clearly bimodal, 
with peaks at 2.35 and 3.10 mag, 
and illustrates the
sensitivity of $V-K$ colours to
metallicity. The bottom panel
shows the colour histogram in
the less sensitive, but most commonly
used $V-I$ colours. Again the colour
distribution is bimodal - but not so well
delineated as the $V-K$ indices - 
with peaks at 0.95 and 1.18 mag.

In the bottom-left panel of Figure~\ref{fig:galform.1} 
we have plotted the observed $V-I$ colour distribution
of NGC~4472 GCs of \citeANP{Larsen01} (2001), 
taken from three \HST\ pointings.
These data have been scaled to the total
numbers of GCs produced in our model.
Comparing the model $V-I$ distributions
with data of limited spatial extent (\eg \HST)
is not ideal, since the relative
number of red to blue GCs changes with 
radius (\eg \citeANP{Geisler96} 1996 ) whilst we model
the entire GC system.
However, bearing this in mind we find that 
for our simulated GC system, 
the widths of blue and red modes in
$V-I$ are 0.08 and 0.07 mag respectively, similar to the NGC~4472
data of \citeANP{Larsen01} (2001) which have widths of 0.08 and
0.09 mag for the blue and red modes respectively.

The relatively old ages of the model
GCs and the broad, distinct metallicity 
distributions shown in Figure~\ref{fig:galform.1}
imply that the observed
colour distributions
are primarily driven by metallicity
differences, consistent with our 
discussion in $\S$~\ref{sec:Introduction}.

\subsection{A Low-Luminosity Elliptical}
\label{subsec:ALow-LuminosityElliptical}

\begin{figure}
\epsfysize 3.4truein
\hfil{\epsffile{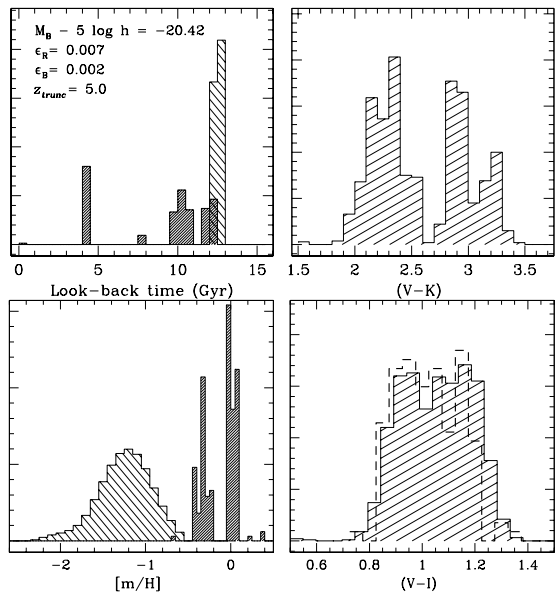}}\hfil
\caption{Age, metallicity and colour 
distributions for the GC system of a low-luminosity 
elliptical.
{\it Top left panel:} 
Age distributions for blue GCs (light shaded histogram)
and red GCs (dark shaded histogram). 
{\it Bottom left panel:} Metallicity 
distributions for blue and red GCs.
{\it Top right:} $V-K$ colours of GCs using the SSP 
models of KFF99.
{\it Bottom right:} $V-I$ colours of GCs derived from
the KFF99 models.
The dashed line indicates the scaled \HST\
data for NGC~4552 from Larsen \etal (2001).}
\label{fig:galform.2}
\end{figure}

In Figure~\ref{fig:galform.2} we show our results for
another GC system, this time for a lower 
luminosity elliptical galaxy, 
which has $M_B$ -- 5 log $h$ = --20.42.
The age and metallicity distributions of the blue
GCs in this galaxy are very similar to those
shown for the more luminous elliptical
in Figure~\ref{fig:galform.1}. 
The blue GCs span the range 
--2.5 $\leq$ [m/H] $\leq$ --0.6, with a mean
of [m/H] $\sim$ --1.2.
Indeed, for the majority of our simulated galaxies, 
variations in the age and metallicity  
distributions of the blue GCs are rather small.
The principal differences in the distributions of the
blue GCs is in the range of their metallicities
for galaxies of different luminosity.
We find that the most luminous galaxies 
possess blue GCs which reach 
to both slightly higher, and in particular, lower metallicities
than galaxies of lower mass.

The distribution of metallicity for the red GCs 
in the low-luminosity elliptical is discontinuous, 
showing intermittent bursts of star formation. 
Indeed, the galaxy in question in Figure~\ref{fig:galform.2}
has undergone 12 major mergers, a factor of $\sim$ 6
fewer than its more massive counterpart.
However, despite this much more modest merger
history, the width and position
of the red GC 'peak' is surprisingly similar
to that of our fiducial galaxy. It is this trend,
in that lower-luminosity ellipticals undergo
fewer major mergers than more luminous galaxies, 
which is the origin of the
scatter in the mean colours of the red GCs
found in Section~\ref{subsec:MeanColours}.

The age distribution for the red GCs in 
Figure~\ref{fig:galform.2} is somewhat different
to that of our luminous fiducial galaxy. 
The distribution of the GC ages, corresponding
to the look-back time of the galaxy's
major mergers, extends to much younger ages. In fact, 
this galaxy has undergone a small merger 
coupled with GC formation recently (400 Myr ago)
with a somewhat stronger merger occurring $\sim$ 4 \Gyr\
ago producing $\sim$ 200 GCs (see next section)
The existence of the very young GCs is reflected in the 
$V-I$ colours of the GCs shown in the bottom-right panel of 
Figure~\ref{fig:galform.2}, which show a population
with $V-I \sim$ 0.5.

The $\sim$ 4 \Gyr\ ages of the 'red' GCs have the  
effect of slightly 'blueing' their $V-I$ colours, 
thereby yielding a broad, top-hat distribution as 
opposed to any obvious bimodality. 
However, this is actually
a fairly rare occurrence in the model and only 1\%
of galaxies actually appear unimodal due to a
spread in ages within the red GC sub-populations 
(see Sec~\ref{subsec:Bimodality}).
The $V-K$ colours (top-right panel) of the GCs again show their 
insensitivity to age differences for all but the very
youngest clusters.

In general we find that, for lower luminosity galaxies, 
usually in lower circular velocity haloes, star formation 
in bursts occurs at successively later times. 
This reflects a generic result of 
semi-analytic models such as \Galform, in that
more massive galaxies in cluster-type environments 
form their stars earlier than less massive galaxies
in the field (we look more closely at this result
in $\S$~\ref{subsec:AgesoftheGlobularClusters}).

In Figure~\ref{fig:galform.2}, we have 
over-plotted \HST\ data from \citeANP{Larsen01}
(2001) for NGC~4552, an E0 galaxy with 
$M_B$ = --20.47, again scaled to the numbers in our
model galaxy GC system. Whilst it is of similar luminosity
to our model galaxy, we have primarily selected these galaxy
data because its bimodality is $not$ clearly delineated, 
providing reasonable agreement with our model
GC $V-I$ colours.
We emphasise that the majority of our simulated 
GC systems, for all galaxy luminosities,
appear clearly bimodal in $V-I$, when the entire
GC system is considered.

\section{Global Properties of the Globular Cluster systems}
\label{sec:GlobalPropertiesoftheGlobularClustersystems}

\subsection{Specific Frequency}
\label{subsec:SpecificFrequency}

A commonly measured property of a galaxy's GC system 
is its total number of GCs. When
normalised to the luminosity
of a 'typical' dwarf galaxy 
($M_V$ = --15; \citeANP{Harris81} 1981)
this becomes the specific frequency:

\begin{equation}
\label{eq:sn}
S_N = N_{\rm GC} \times 10^{0.4(M_V + 15)}
\end{equation}

\noindent where $N_{\rm GC}$ is the total number
of GCs associated with the galaxy in question
with magnitude $M_V$.
Since $\SN$ relates the total
number of GCs to the luminous mass of its
parent galaxy, it is an indication of the efficiency
of a given galaxy at forming GCs, and 
also offers an insight into the 
possible r$\hat{\rm o}$le
of environment upon the properties of
GC systems.

\begin{figure}
\epsfysize 3.4truein
\hfil{\epsffile{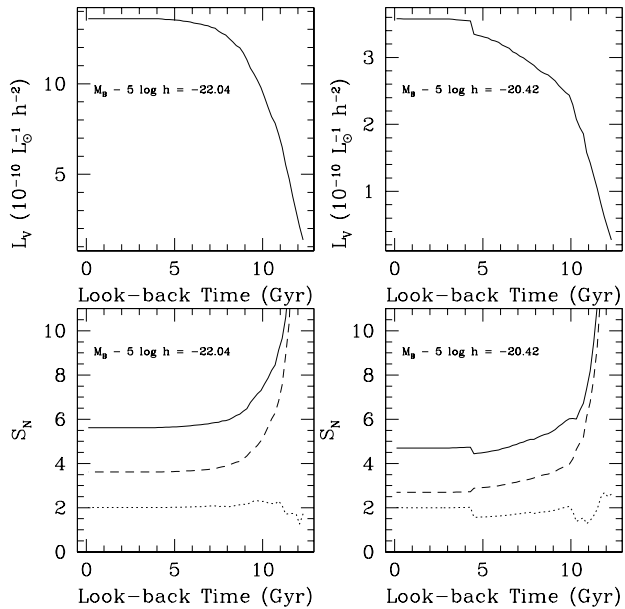}}\hfil
\caption{The evolution of specific frequency 
with look-back time for two model ellipticals. 
{\it Top left:} the luminosity growth of 
our fiducial galaxy. {\it Bottom left:} $\SN$
of the red GCs (dotted line), blue GCs (dashed line)
and total GC system (solid line).
{\it Top right:} the luminosity growth of 
the elliptical with $M_B$ -- 5 log $h$ = --20.42. 
{\it Bottom right:} $\SN$ of the red GCs (dotted line), 
blue GCs (dashed line) and total GC system (solid line).
The 'kinks' in the two right-hand panels occurring
at $\sim$ 4 Gyr correspond to a major merger.}
\label{fig:sn.z}
\end{figure}

We show the luminosity growth 
and the corresponding evolution
of $\SN$ of our fiducial galaxy, and
it low-luminosity counterpart 
in Figure~\ref{fig:sn.z}.
The fiducial elliptical shows
a very smooth increase in its luminosity, 
with some 90\% of its stellar component
formed prior to $z \sim$ 1.0 (a look-back
time of $\sim$ 8 Gyr in our assumed cosmology).
Consequently, the evolution of the $\SN$
in this galaxy's GC systems is also 
very smooth. The $\SN$ of the blue GCs 
is initially very high at early times,
and subsequently declines, taking on a constant value 
at  $z \sim$ 0.4. This occurs because
the blue GCs are all created prior to $z =$ 5, 
whereas the bulk of the galaxy stars form
somewhat later. In contrast, the $\SN$ of the 
red GCs in our fiducial galaxy show very little 
evolution. The creation of red GCs in mergers
prior to $z \sim$ 0.4, is effectively offset
by the formation of field stars, thereby producing
a near constant $\SN$. The observed decline in the 
{\it total} $\SN$ at early times is entirely 
due to the contribution from the blue GCs.

The luminosity growth of the lower-luminosity elliptical
is somewhat different. There is a sudden increase -- a 'kink'--
which occurs at $\sim$ 4 Gyr, indicating that 
the luminous mass of the galaxy increases by
$\sim$ 6\%. 
This kink corresponds to a major merger which this 
galaxy undergoes, leading to the
formation of field-stars and $\sim$ 200 red GCs (see top-left panel of 
Figure~\ref{fig:galform.2}).
Significantly, this late merger has very little
effect upon the total $\SN$ of this elliptical.
Since the formation of the red GCs during this merger 
is accompanied by the formation of some
2 $\times$ 10$^{9} L_{\odot}$ of luminous material, 
the $\SN$ of the red GCs increases by $\sim$ 0.5
whilst the $\SN$ of the blue GCs {\it decreases} by
$\sim$ 0.2. Therefore, the overall effect of this merger
upon the total $\SN$ of the galaxy is only to increase 
$\SN$ by $\sim$ 0.3. Therefore, we find that low-redshift
mergers have no significant effect upon the $\SN$
of ellipticals in our model, reflecting the relatively 
small cold gas fractions involved in mergers at these
epochs.

\begin{figure}
\epsfysize 3.4truein
\hfil{\epsffile{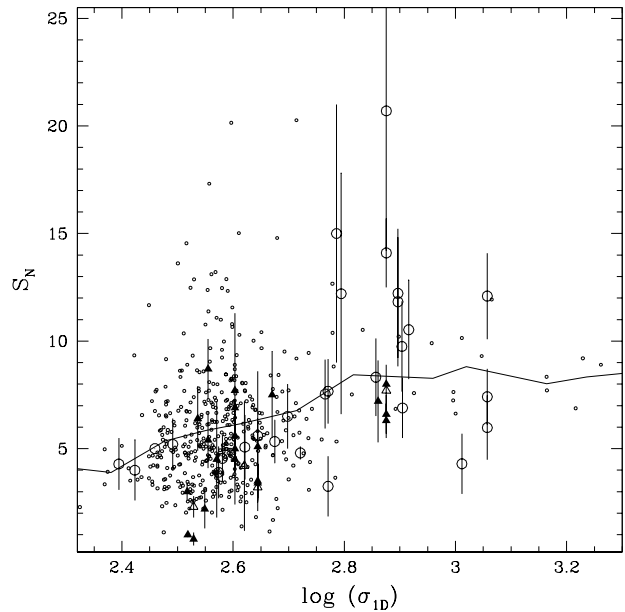}}\hfil
\caption{Dependence of $\SN$ upon 
halo velocity dispersion. 
Small circles indicate the $\SN$ of the
model galaxies, the solid line indicates the mean value
of this $\SN$.
Data are taken from Ashman \& Zepf (1997), 
Blakeslee \etal (1997) and Kissler-Patig (1997).
Open circles represent brightest cluster
galaxies (BCGs) and/or cD galaxies, 
filled triangles represent 'normal' ellipticals 
and open triangles indicate S0s.}
\label{fig:sigma}
\end{figure}

In Figure~\ref{fig:sigma} we show the total $\SN$
of our elliptical galaxies plotted against the 
velocity dispersion of their dark matter haloes.
Here, we use $\sigma_{\rm 1D}$ as a measure of 
cluster mass, obtained from the halo circular velocity, $V_{\rm disc}$, as
discussed in $\S$~\ref{subsubsec:SelectingtheInitialParameters}.
It is important to realise that the 
luminosities of the central dominant
galaxies in \Galform\ (\ie all the galaxies
in this study) are strongly correlated
with $\sigma_{\rm 1D}$. This is a natural
result of semi-analytic galaxy formation
models, where galaxies acquire mass through
merging within dark matter haloes.

In Figure~\ref{fig:sigma}, we have also plotted data 
for 75 early-type galaxies, predominantly taken 
from the compilations of 
\citeANP{KisslerPatig97} (1997) and
\citeANP{AshmanandZepf98} (1998). 
To increase the number of luminous galaxies
in our observational sample, we have also 
included the data of \citeANP{JB97} (1997) and 
\citeANP{Blakeslee97} (1997). These authors obtained 
$\SN$ for 21 ellipticals
in 19 Abell clusters (19 of which 
were brightest cluster galaxies, henceforth BCGs) using
surface-brightness fluctuation techniques. 
Since \citeANP{Blakeslee97} (1997)
measured a 'metric' $\SN$, measured 
within a 40 kpc radius of each galaxy, 
we convert these to global $\SN$ values, 
using the empirical conversion factor
found by \citeANP{Harris98} (1998) of
$\SN$ = 1.3 $\times$ S$_{\rm N}^{40}$.
 
We assign velocity dispersions for the
parent group or cluster for each galaxy
in the \citeANP{KisslerPatig97} (1997) and
\citeANP{AshmanandZepf98} (1998) catalogues, 
based largely upon the galaxy group and cluster
membership given in \citeANP{Rood78} (1978).
In the majority of cases, for groups we obtain 
velocity dispersions directly from \textsc{ned}.
For the \citeANP{Blakeslee97} (1997) sample, 
we use the velocity dispersions given in their
paper (see \citeANP{Harris98} 1998 for a 
similar compilation).

Uncertainties in the total GC population 
of galaxies, which often require large 
extrapolations, lead to uncertainties
in $\SN$ of up to a factor of two for
some galaxies (\eg \citeANP{AshmanandZepf98} 1998).
This uncertainty is comparable to the scatter in 
$\SN$ in the sample used here; therefore searching 
for variations in $\SN$ for ellipticals at a given
luminosity (\eg \citeANP{Harris98} 1998) is difficult.
However, the modelled GC systems shown in Figure~\ref{fig:sigma}
do reflect the observed scatter in $\SN$, but mainly amongst 
elliptical galaxies in haloes with $\sigma_{\rm 1D} <$ 630 \kms.

For galaxies in the centres of clusters, 
$\SN$ is known to correlate
with the dynamical and X-ray mass of the host cluster
(\citeANP{JB97} 1997;
\citeANP{Blakeslee97} 1997; \citeANP{Harris98}
1998). This behaviour is apparent in 
Figure~\ref{fig:sigma}, where many of the observed BCG/cD
galaxies have $\SN$ which are factors of 2--5
higher than their counterparts in lower 
velocity dispersion haloes.
Interestingly, we find that the mean $\SN$ of the model 
ellipticals also increases (by a factor of $\sim$ 2) in going from
the lowest to highest mass haloes, as indicated
by the solid line in the figure. This is, initially, 
a rather surprising result. Since we have adopted $constant$ GC
formation efficiencies (as given in Table~\ref{tab:parameters}), 
one might expect $\SN$ to remain approximately 
constant across all host galaxy magnitudes.

In order to explain the observed correlation
between $\SN$ and cluster mass, 
\citeANP{Blakeslee97} (1997) posited that GCs formed
at early times, and in proportion to the 
total mass available in the 
halo (including dark matter), 
whilst the luminosity growth
of the central galaxy halted at later times, 
yielding the observed correlation of 
$\SN$ with density. 
They interpreted the observed
increase in $\SN$ as a result of the fact that
the BCGs were insufficiently luminous for their 
position in the cluster, \ie the BCGs were under-efficient 
at forming field stars with respect to non-BCG
galaxies, concluding that GCs 
formed with a constant rate per unit mass
(0.7 GCs per 10$^9\Msun$; $\epsilon \sim$ 0.0002).

In a similar vein, \citeANP{Harris98} (1998), 
looked to the hot, X-ray emitting gas associated 
with galaxy clusters to explain the observed 
behaviour of $\SN$. By assuming a constant
GC formation efficiency ($\epsilon \sim$ 0.001),
these authors argued that the baryonic material which
would have gone on to form stars was expelled by
galactic winds during star formation, effectively lowering
the central galaxy's final $stellar$ luminosity 
and effectively increasing its $\SN$.

Developing this idea, \citeANP{McLaughlin99} (1999)
suggested that by considering the initial mass of gas
available to the protogalaxy for forming stars on 
{\it similar spatial scales} to the galaxy starlight, 
the formation efficiency for 
GCs takes a constant value ($\epsilon$ = 0.0026 $\pm$ 0.0005). Moreover, 
he proposed that the ratio $M_{\rm GC}/M_{\rm stars}$ 
does not provide a true measure of $\epsilon$ in the most massive
galaxies (as implied by their larger than average $\SN$), and
that the seemingly greater efficiencies of BCGs in forming 
GCs stems from the fact that $\SN$ only measures the
mass contribution from $stars$, in addition to the fact that 
the $\mathcal{M}/L$ ratios of elliptical galaxies 
vary systematically with their luminosity.

\begin{figure}
\epsfysize 3.4truein
\hfil{\epsffile{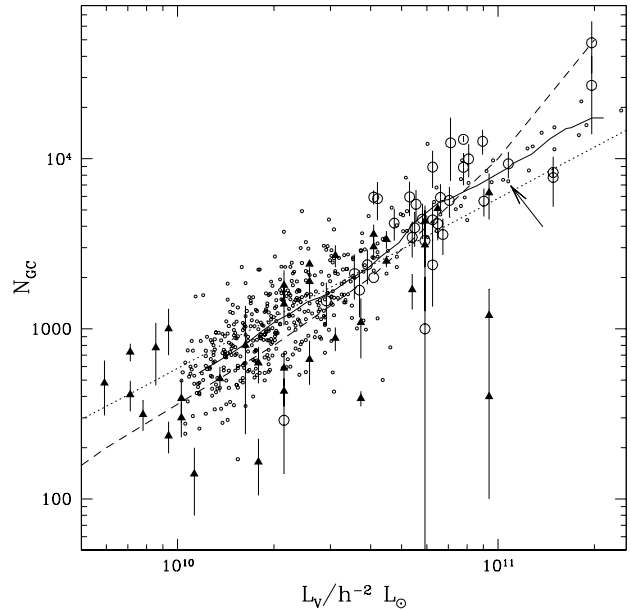}}\hfil
\caption{The total number of GCs formed 
compared to the luminosity of each galaxy.
Small circles indicate the simulated GC systems of 
450 galaxies. The dotted line indicates
$\SN$ = 5 (\ie $N_{\rm GC} \propto
L^{1.0}$). The dashed line is the relation
found by McLaughlin (1999), described in the text.
The solid line indicates the mean value of
$N_{\rm GC}$ for our model elliptical galaxies, the 
arrow indicates the position of our fiducial
model galaxy.
Data are taken from Ashman \& Zepf (1997), 
Blakeslee \etal (1997) and Kissler-Patig (1997).
Open circles represent brightest cluster galaxies 
and/or cD galaxies, filled triangles represent 
'normal' ellipticals and open triangles 
indicate S0s.}
\label{fig:sn.1}
\end{figure}

\begin{figure}
\epsfysize 3.4truein
\hfil{\epsffile{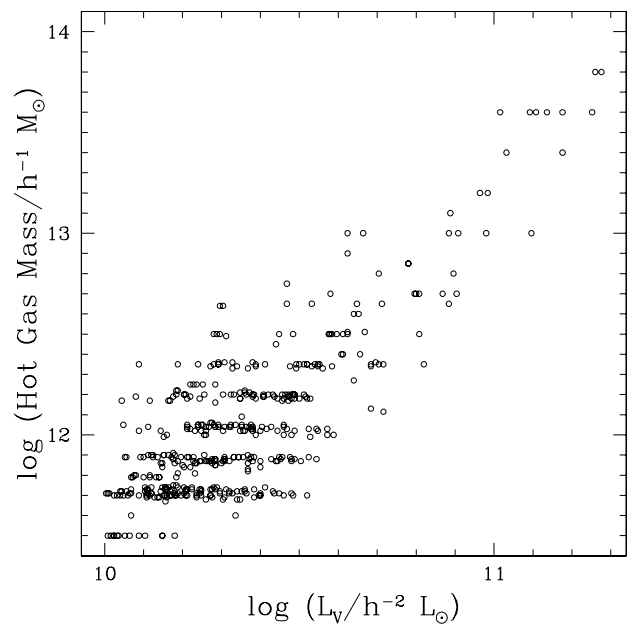}}\hfil
\caption{The total mass of hot gas 
associated with our model galaxy haloes.} 
\label{fig:hotgas}
\end{figure}

Since we explicitly know the hot gas fractions present 
in each galaxy's halo, we can explore these ideas within our 
model. We remind the reader that we have included no dynamical 
destruction of the GCs in our model, and therefore our 
definition of 'efficiency' (equation~\ref{eq:efficiency})
should only be regarded as a true formation efficiency
under this assumption.

\citeANP{McLaughlin99} (1999), defined the observed GC formation
efficiency to include all the hot gas ($M_{\rm hot}$) detected on
the same spatial scale as the galaxy starlight:

\begin{equation}
\label{eq:efficiency.1}
\epsilon \equiv \frac{M_{\rm GC}}{M_{\rm stars} + M_{\rm hot}} \equiv \frac{M_{\rm GC}^{\rm init}}{M_{\rm gas}^{\rm init}}
\end{equation}

\noindent where $M_{\rm GC}^{\rm init}$ is the initial GC
population (before any dynamical processes have acted upon
the GCs) and $M_{\rm gas}^{\rm init}$ the initial gas supply 
available to the protogalaxy.
Under the assumption of a constant GC formation efficiency,
\citeANP{McLaughlin99} (1999) found that the total number of 
GCs ($N_{\rm GC}$), scaled thus:

\begin{equation}
\label{eq:ngc}
N_{\rm GC} \propto L_{V, \rm gal}^{1.3} \left(\frac{M_{\rm hot}}{{M_{\rm stars}}}\right)
\end{equation}

\noindent where $L_{V, \rm gal}$ is the V-band luminosity
of the host galaxy. The scaling $L_{V, \rm gal}^{1.3}$
comes from noting that the $\mathcal{M}/L$ ratios of 
ellipticals systematically increase with their luminosity, 
a result similar to those found previously by
\citeANP{Djorgovski92} (1992) and \citeANP{Zepf94} (1994).

The scaling given in equation~(\ref{eq:ngc}), is indicated
by the dashed line in Figure~\ref{fig:sn.1}. Here we have
plotted the total number of GCs formed for each model galaxy, against
host galaxy $V$-band luminosity (see figure 8 of
\citeANP{McLaughlin99} 1999 for a similar plot).
Again, we have included the observational data described
previously.

The mean values of $N_{\rm GC}$ for the model ellipticals, 
systematically deviate from the $\SN$ = 5 line (
\ie a scaling of $N_{\rm GC} \propto L_{V, \rm gal}^{1.0}$). 
Whilst the precise location of the model points in
the $y$-axis depends upon our adopted efficiencies (the
position of our fiducial galaxy is indicated in
Figure~\ref{fig:sn.1}), a least-squares fit to the model points 
indicates a scaling of $N_{\rm GC} \propto L_{V, \rm gal}^{1.25}$. 
The dependence of $N_{\rm GC}$ upon $L$ for our model galaxies
is similar to the scaling suggested by 
\citeANP{McLaughlin99} (1999; c.f. equation~\ref{eq:ngc}), to
account for the observed increase in $\mathcal{M}/L$ ratios of elliptical
galaxies with luminosity. Since we have assumed $\epsilon$ =
constant, and we have not included the model hot gas component in the
calculation of $N_{\rm GC}$, the increase in model $\SN$ can only
be accounted for by an increasing $\mathcal{M}/L$ ratio of our
elliptical galaxies with increasing luminosity.
As discussed by \citeANP{Benson00} (2000; see their figure
8), the \B-band $\mathcal{M}/L$ ratios of galaxy haloes in \Galform\ are
a function of halo mass. For the halo mass range considered
in this study (1.0$\times 10^{13} h^{-1} \Msun \leq M_{\rm halo} \leq
1.3 \times 10^{15} h^{-1} \Msun$), the $\mathcal{M}/L$ ratio
of galaxy haloes is an increasing function of $L$.
This occurs because at $M_{\rm halo} > 10^{12} h^{-1} \Msun$, 
gas cooling is increasingly inefficient over the halo lifetime, 
thereby inhibiting galaxy formation (the most efficient 
value of $M_{\rm halo}$ is $\sim 10^{12} h^{-1} \Msun$, below these
masses galaxy formation is inhibited by star formation feedback
-- see \citeANP{Benson00} 2000).

The mean line of the model galaxies in Figure~\ref{fig:sn.1}
scales in the correct sense with the observational data, 
and the \citeANP{McLaughlin99} (1999) relation. However, the
model is still inconsistent with the highest $\SN$ galaxies
(although it may be argued that the scatter is large enough to
be consistent). 
To account for this remaining discrepancy, \citeANP{McLaughlin99} (1999)
included the hot gas component in the calculation of $N_{\rm
GC}$, as indicated by equation~\ref{eq:ngc}.

We show the behaviour of the model hot gas component 
in Figure~\ref{fig:hotgas}. Here, 'hot gas' refers
to all the baryonic material in the halo which is not in the 
form of cold gas or stars.
Figure~\ref{fig:hotgas} indicates that the 
most massive central cluster galaxies are able to expel and
virialize large quantities of gas at early
times, due to a combination of feedback from
star formation, and long gas-cooling time-scales.
This picture is supported by the observational
studies of \citeANP{Harris98} (1998) and \citeANP{McLaughlin99}
(1999).
Moreover, it is also clear that the halo hot
gas component of our model ellipticals increases much $too$ 
strongly (by a factor of $\sim$ 100) to account for 
the differences shown in Figure~\ref{fig:sn.1}.

However, consideration of this apparent inconsistency with the
\citeANP{McLaughlin99} (1999) study must be accompanied with
an important caveat. 
The X-ray gas considered by \citeANP{McLaughlin99} was measured
on $<$ 100 \kpc\ scales (\ie on scales where the galaxian surface 
brightness can be measured), whereas the hot gas component 
in \Galform\ is associated with the entire galaxy halo (\ie on
$\sim$ 500 \kpc\ scales). Associating some core radius of hot 
gas in our model with galaxy luminosity is not particularly
meaningful, since in our model the entire hot gas component 
is associated with the star formation in
the central galaxy. Our scheme is more in keeping with
the interpretations of \citeANP{West95} (1995),
\citeANP{Blakeslee97} (1997) and \citeANP{Harris98} (1998), 
in that the 'high $\SN$' problem is associated with  
entire haloes (\eg clusters), rather than on the more
local scales as suggested by \citeANP{McLaughlin99} (1999). 
These interesting issues will be examined in greater
detail in a future paper.

\subsection{Mean Colours}
\label{subsec:MeanColours}

Since the initial study by \citeANP{vandenBergh75}
(1975), the subject of whether or not the mean 
metallicities of GC systems (and their sub-populations)
correlate with host galaxy luminosity 
has been a controversial one (\eg \citeANP{Brodie90} 1990;
\citeANP{Ashman93a} 1993; 
\citeANP{Secker95} 1995; \citeANP{Durrell96} 1996;
\citeANP{Forbes96a} 1996; \citeANP{Kundu01} 2001; 
\citeANP{Forbes01} 2001; \citeANP{Larsen01} 2001).
Since in our model, we have information
about the stellar populations of both the
GCs and their host galaxies, we are in a 
position to look for any such correlations
between these two properties.

\begin{figure}
\epsfysize 3.4truein
\hfil{\epsffile{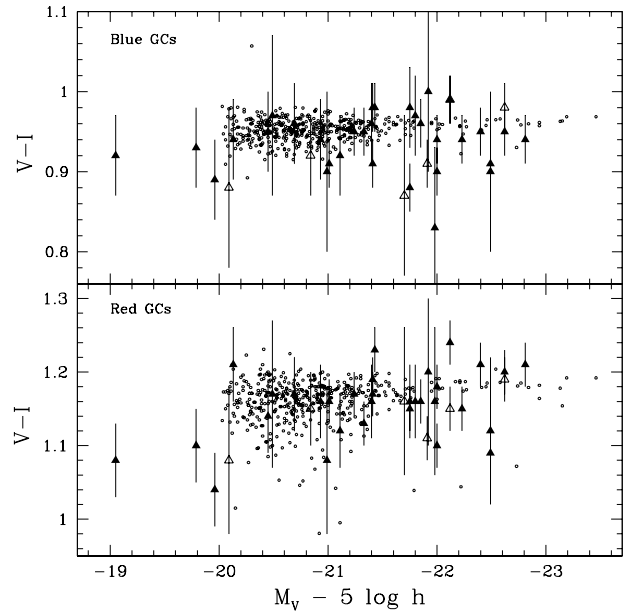}}\hfil
\caption{The mean colours of the GC systems
with host galaxy magnitude. {\it Top panel:}
blue GCs, {\it bottom panel:} red GCs.
Data points with error-bars
taken from the compilation of Forbes \& Forte (2001). 
Open triangles represent the GC systems of
S0 galaxies, filled triangles represent galaxy types earlier
than S0, }
\label{fig:meanVI}
\end{figure}

In Figure~\ref{fig:meanVI} we have plotted
the mean $V-I$ colours of the red and blue GCs
predicted by our model against the 
absolute \V-band magnitude of the host galaxy. 
The data-points are taken from the compilation of
\citeANP{Forbes01} (2001), who have placed the 
GC colour distributions of 28 galaxies
onto a common $V-I$ system. To facilitate
a fair comparison, we have omitted
all the galaxies from the \citeANP{Forbes01} (2001)
compilation with morphological type later than S0.

For both the red and the blue GCs, our
model points lie comfortably within 
the observational uncertainties of these
data. The large scatter in our model
at the low-luminosity end arises partly because of the 
occurrence of late bursts of GC formation, where
age effects become important in determining the colours
of the GCs.

Our model predicts an essentially flat 
relation between host galaxy magnitude 
and mean (although
there is an increase in both sub-populations 
at the 1-$\sigma$ level). 
\citeANP{Forbes01} (2001) found 
that the mean $V-I$ colours of the red GCs increased
with both host galaxy luminosity and velocity
dispersion.
These authors argued that such a correlation 
was indicative of a close connection between the red GCs 
and the bulk stellar content of their host galaxy,
and that this connection primarily supported the multi-phase collapse model
of \citeANP{Forbes97} (1997). 
Interestingly, \citeANP{Kundu01} (2001) also found evidence for 
such a correlation (from the thesis work of
A. Kundu, from which part of the \citeANP{Forbes01} sample was
derived), however these authors did not ascribe any 
particular significance to their findings, and argued that 
such a correlation did not support any one particular GC
formation model.

At face value, the flat colour-magnitude gradients
of the model GC systems indicate that 
either {\it (i)} the GCs do not 
'know' about their host galaxy, implying that the GCs do not 
share a common  chemical enrichment history  
with the galaxy or {\it (ii)} all the galaxies
in our sample are enriched to very similar levels.

We know the former scenario not to be the case,
since in our scheme the formation of red GCs
is tied to the star formation 
which is associated with major mergers. 
Any subsequent increase in the metallicity of these 
bursts would lead to enrichment of the GCs, and 
therefore be reflected in their colours.
The situation for the blue GCs is not so clear, 
since we truncate their formation at the early stages 
of the formation of their host galaxy.
However, it seems reasonable to expect 
that enrichment in these GCs will have preceeded to 
similar levels irrespective of the final mass 
of the galaxy, leaving little or no correlation
between their mean colours and host galaxy 
luminosity. 

\begin{figure}
\epsfysize 3.4truein
\hfil{\epsffile{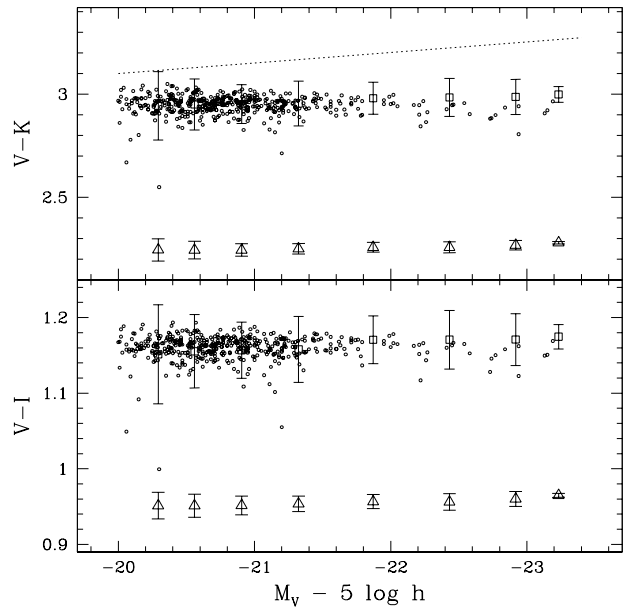}}\hfil
\caption{The mean colours of the GCs and their
host galaxies, versus parent galaxy magnitude.
{\it Top panel:}  Mean $V-K$ colours of the red
GCs (squares with 1$\sigma$ error-bars), 
blue GCs (triangles with 1$\sigma$ error-bars) 
and model galaxies (small
circles) versus galaxy $M_V$. The dotted
line indicates the slope of the galaxy colour-magnitude 
relation for Coma galaxies from Bower \etal (1992).
{\it Bottom panel:} Same for $V-I$ colours.}
\label{fig:galaxy}
\end{figure}

To better understand these results, we look 
at the behaviour of the GC mean colours their parent galaxies 
with galaxy magnitude in Figure~\ref{fig:galaxy}.
The mean colours of the galaxy sample
are determined from the mass-weighted
age and metallicity of their stellar populations, however we find that 
by adopting a luminosity-weighting scheme has little 
effect on the colours.
We have not included any corrections for 
aperture in the colours of the galaxies, 
which can have a significant effect on their
mean colours (\eg \citeANP{Kodama98} 1998).
However, since we have no spatial information about
the stellar populations in our sample and we cannot
correct for the presence of radial gradients
which are seen in real galaxies, we simply state 
that the galaxy colours given in Figure~\ref{fig:galaxy} are 
a lower limit.

The figure shows that the mean colours of the red
GCs are generally consistent with the mean galaxy
colours, although for the most luminous
galaxies the GC colours are marginally $redder$
than the galaxies.
The blue GCs are some 0.2 mag bluer in
$V-I$ than their host galaxy, corresponding
to a pure metallicity difference of $\sim$ 1.0 dex.
This is consistent with all studies to date
for which photometry of both the GC sub-populations 
$and$ the host galaxy light have been obtained in the same
observation (NGC~1399 : \citeANP{Ostrov98} 1998, 
NGC~1427 : \citeANP{Forte01} 2001, NGC~3923 : 
\citeANP{Zepf95} 1995, NGC~4472 : \citeANP{Geisler96} 1996;
\citeANP{Lee00} 2000), as noted by \citeANP{Forbes01} (2001).

Figure~\ref{fig:galaxy} also highlights another
important issue: the mean model galaxy colours show no increase
with magnitude. Indeed they exhibit very similar behaviour 
to that of the GCs. Compare this to the colour-magnitude ({\it C-M}) 
relation found for Coma cluster galaxies by
\citeANP{Bower92} (1992), indicated by the
dotted line in the top panel of Figure~\ref{fig:galaxy}.
The lack of a significant colour gradient 
amongst the red GCs is seemingly a manifestation 
of the inability of the semi-analytic models 
(including \Galform) to naturally
reproduce the observed {\it C-M} relation
of galaxies.
Whilst at the lowest luminosities
($M_V$ -- 5 log $h~\geq$ --19.0), the slope
of the {\it C-M} relation is recovered (\eg
see figures 10 \& 13 of \citeANP{Cole00} 2000), 
above this the mean galaxian metallicities
reach a peak beyond which chemical enrichment does
not proceed further. This precludes any increase 
in the mean galaxy colour, thereby giving 
rise to a flat (or even negatively sloped) 
{\it C-M} relation. 
 
Such behaviour can be remedied by
assuming extremely strong feedback, 
as was shown by \citeANP{Kauffmann98} (1998).
However, adopting very large values for
the feedback efficiency yields
unphysical results in other predictions
of \Galform\ (see \citeANP{Cole00} 2000
for details).
This failure of the semi-analytic models
probably belies our lack of understanding
of the complex processes of star formation, 
and is beyond the scope of this paper.

\subsection{Bimodality}
\label{subsec:Bimodality}

One of the motivations of this study is
to reproduce the observed bimodality
of GC systems, and in doing so attempt
to understand their physical origins.
In $\S$~\ref{subsubsec:SelectingtheInitialParameters},
we found that it was necessary to truncate 
the blue GC formation at early times (\ie $z$ = 5)
in order to reproduce GC colour
distributions similar to those which are observed.

\begin{figure}
\epsfysize 3.8truein
\hfil{\epsffile{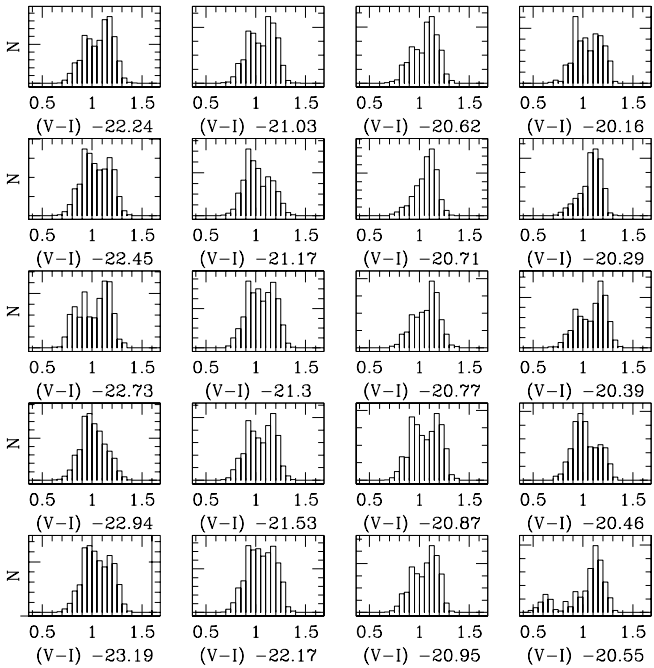}}\hfil
\caption{The $V-I$ colour distributions for a selection
of simulated GC systems.
The values at the bottom of each plot indicate
the \V-band magnitude ($M_V$ -- 5 log $h$) of each galaxy.
The distributions show a wide variety of morphologies, 
from principally blue-peaked, to bimodal, to strongly red-peaked.} 
\label{fig:bimodal}
\end{figure}

In spite of this enforced truncation of the blue 
GCs, the colour distributions of the entire GC 
systems show an appreciable variation in their 
morphologies. We show a selection of $V-I$ colour distributions 
for our simulated GC systems in Figure~\ref{fig:bimodal}. 
The galaxy GC systems are ordered
by descending host galaxy magnitude
(from the bottom left to top right in the figure).
The colour distributions range from bimodal 
(\eg the elliptical with $M_V$ -- 5 log $h$ = --23.19), 
to predominantly blue-peaked (\eg $M_V$ -- 5 log $h$ = --22.94)
and predominantly red-peaked (\eg $M_V$ -- 5 log $h$ = --20.29).
The elliptical with $M_V$ -- 5 log $h$ = --20.55 shows
a number of GCs with 0.5 $\leq  V-I \leq 0.8$.  
This is the result of several recent mergers between 
this galaxy and satellites, which 
gives rise to significant star formation 
accompanied by GC formation.

In order to make a fair comparison between
the GC colour distributions predicted by our
model, and those observed, we
must restrict our analysis of bimodality to
some sub-sample of the GC population.
Taking the study of \citeANP{Gebhardt99} (1999) 
as being representative
of a 'typical' \HST\ observation of a 
GC system, we return to our bimodal
prototype galaxy NGC~4472. From two \HST\ pointings, these
authors identified 486 GCs out of
a total population of $\sim$ 6400 GCs
\cite{Geisler96}, thus sampling
approximately 8\% of the total GC population
of this galaxy. Whilst subsequent studies
have gone deeper for this galaxy, and/or have increased
pointings 
(\eg \citeANP{Puzia99} 1999; 
\citeANP{Lee00} 2000; \citeANP{Larsen01} 2001), 
we take the data of \citeANP{Gebhardt99} (1999)
as a lower limit.

To simulate these observations, 
we take a random distribution of GCs corresponding
$\sim$ 8\% of each GC system.
We then assign each of these 
GCs a luminosity, drawn randomly 
from an observed luminosity
function. Specifically, the luminosity
function adopted is that of NGC~4472, 
taken from \citeANP{Puzia99} (1999).
Finally, we assign photometric errors to
each GCs colour (see Appendix A in \citeANP{Harris01} 2001):

\begin{equation}
\label{eq:error}
\sigma_{\rm phot} \simeq 2.5 \times {\rm log}\left(1+\frac{1}{(S/N)}\right)
\end{equation}

\noindent where $S/N$ is the signal-to-noise and $\sigma_{\rm phot}$
the photometric uncertainty.

\begin{figure}
\epsfysize 3.2truein
\hfil{\epsffile{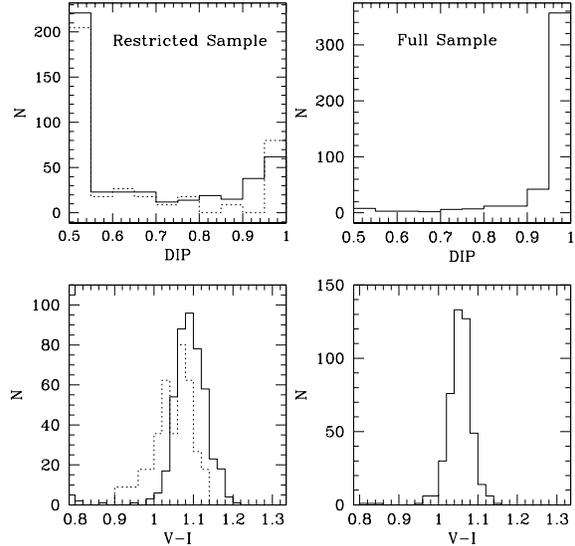}}\hfil
\caption{The DIP statistic and mean colours of our model
GC systems.
{\it Top left:} DIP statistic for our simulated
\HST\ observations (solid lines), compared to the data of 
Gebhardt \& Kissler-Patig (1999; dotted lines).
{\it Top right:} Distribution of the DIP statistic for the 
complete model GC systems of each elliptical.
{\it Bottom right:}  mean $V-I$ colours of the 
complete model GC systems and {\it bottom left:} mean $V-I$
colours of our simulated \HST\ observations (solid lines), 
compared to the data of Gebhardt \& Kissler-Patig 
(1999; dotted lines).
}
\label{fig:dip}
\end{figure}

To look for the presence of multiple populations
in restricted datasets, statistical tests such as
the KMM test \cite{Ashman94} and the 'DIP statistic'
(\eg \citeANP{Hartigan85} 1985; \citeANP{Gebhardt99} 1999)
are often employed. The success of such tests
necessarily depend, among other things, upon
the size of the dataset considered.
To place our results on a more quantitative
footing, we have measured the DIP statistic and
mean colours of the colour distributions of our full 
GC sample and simulated \HST\ observations. The DIP statistic 
is a non-parametric test whose null hypothesis is that a distribution
is unimodal (see \citeANP{Gebhardt99} 1999). A value near 0.5 indicates
likely 'uni-modality', whereas 1.0 indicates
the presence of multiple sub-populations
(not necessarily bimodality, although for our model 
this is assumed to be the case).

We show the results of these
statistical measures in Figure~\ref{fig:dip}.
The left-hand panels in the
figure show the results for our
simulated \HST\ observations, compared to 
the scaled data from \citeANP{Gebhardt99} (1999).
The right-hand panels in the indicate the results 
for our unrestricted model GC systems.

Figure~\ref{fig:dip} indicates that both the model
\HST\ observations and data are distributed in a similar
manner, with a large 'spike' at 0.5 (uni-modality), 
and a somewhat smaller one at 1.0 (multi-modality), 
coupled with a relatively flat distribution of DIP values
in between. The model predicts a slightly
higher fraction of GC systems lying at 
DIP 0.8 $\sim$ 0.9 than is the case 
for the \HST\ data.
\citeANP{Gebhardt99} (1999) separated their data into
three groups, those galaxies which showed
a clear deviation from uni-modality, those
which had DIP values higher than 0.5 at greater
than 1$\sigma$ significance and galaxies which
appeared to have unimodal GC systems.
We have performed the same division on our data, using the
same definitions as \citeANP{Gebhardt99} (1999). We
define 'certain' as systems which are deemed bimodal
by the DIP statistic, 'likely' as likely bimodal, 
and 'unimodal' as those GC systems which do not appear 
bimodal.

From our 450 GC systems, we
obtain a relative fraction (certain : likely : unimodal) 
of 86 : 118 : 246, or 19\% : 26\% : 55\%.
These values are similar to the \citeANP{Gebhardt99}
results, which comprise of 21\% : 26\% : 53\%.
However, we find that these values are
extremely sensitive to the model 'photometric
errors' which we adopt. 
Increasing the mean uncertainties by only 0.01 mag 
in $V-I$, yields a ratio of 16\% : 25\% : 59\% (\ie more
bimodal systems are hidden by photometric errors),
likewise, reducing the uncertainties by 0.01 mag
results in a 24\% : 30\% : 45\% split.
This sensitivity to photometric errors 
is probably the origin of the different measures of 
bimodality for the same GC systems between 
different studies.

However, by obtaining the DIP statistic for our unrestricted GC 
systems (right-hand panels in Figure~\ref{fig:dip}),
we find that 93\% are intrinsically
bimodal in $V-I$. 
Of the remainder, 4\% fall into the 'likely' category
and 3\% are not bimodal.\footnote{The DIP statistic is somewhat
more conservative than some other statistical methods for 
detecting bimodality in datasets. Applying the homoscedastic
(a common covariance) KMM test to our results indicates that 
96\% of the model GC systems are not unimodal at high confidence.} 
The majority of
these latter 32 galaxies have less than 20 GCs in their
total GC system, and small-number statistics
become the dominant factor in determining 
the presence of sub-populations in these systems.
Five galaxies (1\% of the total) have undergone 
recent mergers coupled with the formation of
GCs, which conspire to superimpose the red and blue 
GC sub-populations upon each other, giving apparently unimodal
distributions in $V-I$. When examined
in $V-K$ colours, three of these
systems appear bimodal (the other two
are effectively present-day mergers, and therefore
even the $V-K$ colours are extremely blue).
Thus, in our model, the GC systems
of nearly all large elliptical galaxies
are intrinsically bimodal in broad-band 
colours, which is detectable 
if the total GC population 
can be 'observed'.
It then follows that since nearly every 
non-detection of bimodality (from an intrinsically
bimodal parent distribution)
is attributable to photometric
errors and/or small-number statistics, 
(irrespective of whether our model
is correct or not), any conclusions relating to the 
colour distributions of GC systems obtained 
from observations of only limited depth and 
spatial extent should be made cautiously.

The bottom-left panel of
Figure~\ref{fig:dip} compares the
biweight estimate of the position (sample
mean) of the $V-I$ peaks of the
simulated \HST\ observations and 
observational data.
The position of the peaks are 
similar, with $V-I$ = 1.09 and  $V-I$ =
1.05 for model and data respectively, whilst 
the distribution in the
model colours is slightly narrower than
that of these data,  
with $\sigma_{\rm rms}$ = 0.049 and 
 $\sigma_{\rm rms}$ = 0.052 respectively (the \HST\
data has been scaled by a factor
of $\sim$ 9 in the figure).
For the unrestricted model GC systems, 
the mean $V-I$ colours is 1.06, with 
a dispersion of 0.035 mag. 
The relatively narrow distribution
of these colours indicates that 
the mean of the GC colour
distributions is fairly insensitive to 
the varied merger histories of the individual
galaxies. 

\subsection{Ages of the Globular Clusters}
\label{subsec:AgesoftheGlobularClusters}

A key observational test of the various
models for the formation of GC systems 
lies in determining their age.
In our model, we know the formation redshift 
of each GC, and therefore we
can look at the age distributions of GCs 
amongst our galaxy sample.

\begin{figure}
\epsfysize 3.4truein
\hfil{\epsffile{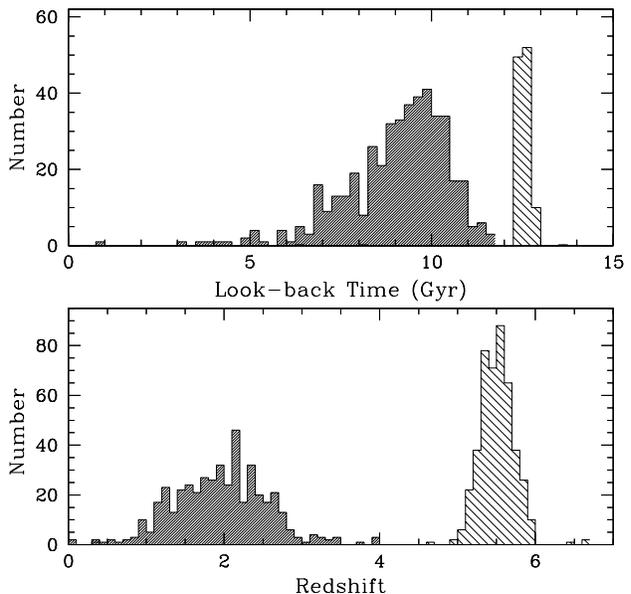}}\hfil
\caption{The distribution in mean ages
for all 450 model GC systems. 
{\it Top panel:} Distribution in look-back
time for the red GCs (dark shaded histogram)
and blue GCs (light shaded histogram). 
The blue GC histogram has been scaled down
by a factor of 6 for display purposes.
{\it Bottom panel:} Distribution in formation 
redshift for the red and blue GCs.}
\label{fig:age.1}
\end{figure}

In Figure~\ref{fig:age.1} we show the
mean age (look-back time) and formation 
redshift distributions for our model
GCs (the cosmological parameters adopted
are given in Table~\ref{tab:parameters}).
Perhaps the most striking feature 
in the figure is the narrow
range of mean ages occupied by the 
blue GCs, at $\sim$ 12.3 \Gyr\ (top panel).
This mean age for the blue GCs
is consistent with the 'favourite' age of Galactic
GCs found by \citeANP{Carretta00} (2000) of 
12.9 $\pm$ 2.9 \Gyr. 
Whilst the onset of blue GC formation begins
as soon as enough gas cools to begin 
forming stars (at $z \sim$ 11; c.f. Figure~\ref{fig:truncate})
the vast majority of blue GCs form at or around our 
truncation redshift.

The red GCs clearly form over a more
extended period, with mean
redshifts of formation from $z \sim$ 4 
to the present day. However, red GCs $begin$ to form 
in small numbers at $z= 9 \sim 10$ during the onset of
PGD merging (e.g Figure~\ref{fig:truncate}).
The mean ages of the red GCs
occurs at 9 \Gyr\ ($z \sim$ 2), although the skewed
nature of the age distribution
means that the peak of red GC formation
actually occurs $\sim$ 10 \Gyr\ ago. 
This gap in the formation redshifts
of the blue and red GCs is key to the
metallicity differences between the GCs.
The material fed-back to the interstellar
medium after the first period of 
GC formation has time to cool and 
form GCs through merging 
at a later epoch.

\begin{figure}
\epsfysize 3.4truein
\hfil{\epsffile{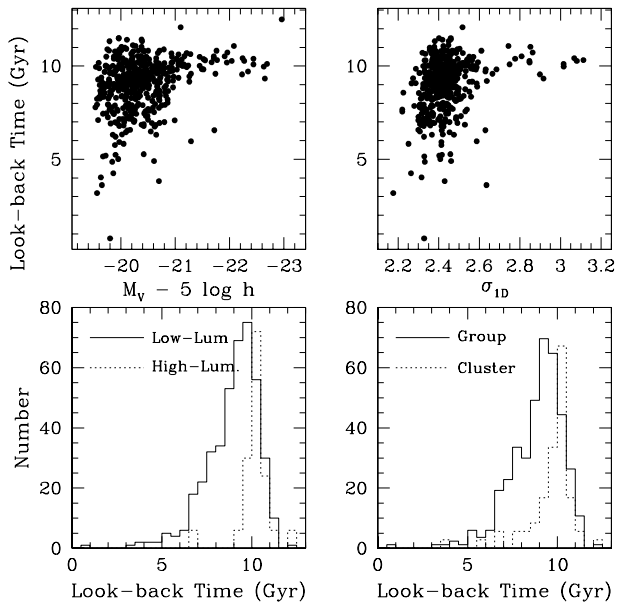}}\hfil
\caption{The mean ages of the red GCs as
a function of host galaxy properties. 
{\it Top left panel:} Mean ages of the red GCs
as a function of host galaxy absolute magnitude.
{\it Bottom left panel:} Histogram of the mean GC ages, 
separated into high and low-luminosity
ellipticals. The histogram for the high-luminosity
galaxies has been scaled by a factor of 3 for
display purposes.
{\it Top right panel:} Mean ages of the red GCs
as a function of halo velocity dispersion.
{\it Bottom right panel:} Histogram of the mean red GC ages, 
separated into field/group and cluster environments.
The histogram for the field/group
galaxies has been scaled by a factor of 5 for
display purposes.}
\label{fig:age.2}
\end{figure}

The short duration of the blue GC formation 
makes comparisons of their ages with 
other galaxy properties difficult.
Moreover, the mean ages of the blue GCs are clearly 
sensitive to the exact redshift of their truncation.
However, this is not the case
for the red GCs, which form over
a significant fraction of a Hubble time.
In Figure~\ref{fig:age.2}, 
we show the mean ages of the red GCs as a 
function of their host galaxy magnitude 
and halo velocity dispersion. 
The behaviour of mean ages of the red GCs
with host galaxy luminosity and halo velocity
dispersion are very similar. The highest 
luminosity galaxies, and those galaxies in 
the highest $\sigma_{\rm 1D}$ haloes both give rise
to red GCs with ages $\sim$ 10 \Gyr, with small 
scatter. In proceeding to lower 
luminosities (and lower $\sigma_{\rm 1D}$), 
this scatter in mean age increases sharply, with
a tendency for the GCs to possess younger ages.

In the bottom panels of the figure, we
show histograms of the mean GC ages 
separated by their host galaxy properties, 
according to our definitions given
in $\S$~\ref{subsubsec:TheGalaxySample}.
We find that the red GCs associated with 
low-luminosity ellipticals have mean ages
of 8.8 \Gyr\ ($\sigma_{\rm rms}$ = 1.4 \Gyr), whereas for the 
high-luminosity galaxies the mean age of the 
GC systems is 10.0 \Gyr\ ($\sigma_{\rm rms}$ = 0.8 \Gyr).
Similarly, the GC systems associated with
ellipticals in the field and/or groups
have mean ages of 7.9 \Gyr\ ($\sigma_{\rm rms}$ = 1.7 \Gyr), 
whilst those in clusters have mean ages of
9.2 \Gyr\ ($\sigma_{\rm rms}$ = 1.3 \Gyr).
Clearly, these values will change depending upon
the exact nature of the definitions of 'low-luminosity elliptical' or 
the field/group and cluster environments, however the general 
trends will remain.



\section{Discussion}
\label{sec:Discussion}

There now seems little doubt that GCs are formed
during the merging and interaction of galaxies
which involve star formation (\eg \citeANP{Schweizer01} 2001).
In this regard a metal-rich GC component 
arises naturally in a hierarchical model 
such as \Galform, which predicts many such mergers 
over a broad distribution of redshifts. 
However, to what $degree$ these merger-formed GCs contribute
to the red peak associated with the GC populations of
elliptical galaxies is uncertain.

In our scheme, continuous gaseous merging produces the
entire red GC sub-population observed in ellipticals, and 
the ages of these GCs directly reflect the redshifts at which their 
host galaxy underwent major mergers involving star formation. 
In this way, the red GCs represent a direct probe of the 
dynamical and stellar assembly of their host galaxies.
As shown previously, the red GC sub-populations
of our model ellipticals possess significant age structure 
in any given galaxy. Moreover, the scatter in this age structure is 
predicted to increase with decreasing host-galaxy
luminosity and decreasing dark matter halo velocity 
dispersion, reflecting 
the occurrence of later star formation in these
less-massive haloes. This is a principal result
of this study, and provides a method of testing our simulations
against other scenarios for forming the GC systems of 
luminous elliptical galaxies.

In the {\it in-situ} collapse model of 
\citeANP{Forbes97} (1997), two phases 
of galaxy collapse give rise to two GC populations 
enriched to significantly different levels.
The first collapse produces metal-poor GCs with high
efficiency with respect field stars, followed at some
later time by a secondary collapse where the majority 
of the galaxy stars and metal-rich GCs are created.
In common with our model, the metal-poor 
GCs are predicted to be several Gyr $older$ than
the red GCs, with the exact age difference depending 
upon the delay between the two collapses. If this
delay has a characteristic time-scale, such as that for the
onset of type-I supernovae, then the age difference between
the GC sub-populations is expected to be similar for galaxies
irrespective of galaxy luminosity and halo velocity dispersion. 
Moreover, the metal-rich GCs are will
be essentially $coeval$, since the time-scale 
for the secondary collapse is expected to be short.

\citeANP{Cote98} (1998) presented a  
model where the metal-poor GCs are captured from the 
surrounding galaxy population via tidal stripping and/or mergers,
and the red GCs are somehow intrinsic to the host elliptical, 
in common with the {\it in-situ} model. These authors
found that by adopting a relatively steep luminosity function of
galaxies in the host galaxy cluster, the bimodality of 
GC metallicities could be reproduced without necessarily 
requiring two phases of star formation. Whilst \citeANP{Cote98} (1998)
do not discuss in detail the possible mechanism for the formation of either
GC sub-population, they do predict that there will exist $no$ 
significant age differences between the GC sub-populations.

In many respects, the model presented here closely
follows the original philosophy behind the Ashman \& Zepf (1992)
merger model. Merging creates metal-rich GCs, whilst the
merger progenitors contribute their own population
of metal-poor GCs to the final GC system.
Since the spiral-spiral mergers of Ashman \& Zepf
(1992) are expected to occur at relatively high redshift, 
it may be argued that the distinction between the 
proto-galactic discs in our model and gaseous 
spirals in the Ashman \& Zepf picture becomes
ill defined. 

However, although it might be tempting to associate these 
PGDs with the precursors of 
todays spiral galaxies, this is $not$ the case.
In the semi-analytic model as presented by \citeANP{Cole00} (2000), 
spiral galaxies form through the accretion of cold gas onto
bulges. Providing no further merging occurs, this gas will settle
and remain in a stellar/gas disc. A large initial bulge mass with
respect to the mass of accreted cold gas will correspond to an 
early-type spiral, and a less-massive bulge 
will correspond to a later-type spiral.
An obvious extension of our model will be to study the GC systems
of late-type galaxies, however the implication is clear:
in our model, the bulges of spirals are formed in
the same manner as those of ellipticals. 
Therefore, {\it the metal-rich GCs in spirals are
associated with the bulge and/or thick disc, and 
the majority of spiral galaxies will have bimodal GC metallicity
distributions.} This notion that the GCs of ellipticals and
spirals have essentially the same origin finds observational
support in the recent study of \citeANP{Forbes01a} (2001). 

In our simulations, the progenitors 
of the blue GCs are exponential, proto-galactic discs, which 
at high redshifts, are largely gaseous. 
For example, at $z \sim 5$, these PGDs 
have mean cold gas-fractions of order 95\%. 
We assume that these PGDs
form, and continue to form, blue GCs (prior to our 
truncation redshift) unless they merge with the
central galaxy in the dark matter halo.
As suggested by \citeANP{Ashman93} (1993), 
it is then not difficult to imagine that any such 
PGDs which failed to merge are now represented 
by dwarf galaxies (\eg the LMC), which have only
managed to form metal-poor, blue GCs.

Observationally, perhaps the best candidates 
for such proto-galactic entities are damped Ly$\alpha$ 
systems (DLAs), high-redshift H\thinspace\textsc{i}
systems with column densities N$_{\rm H\thinspace\textsc{i}} 
\ge 2 \times$ 10$^{20}$ cm$^{-2}$, 
seen in absorption against background QSOs 
(\eg \citeANP{Wolfe88} 1988). \citeANP{Burgarella01}
(2001) have recently discussed the possible relation
between metal-poor GCs and DLAs at high redshift.
\nocite{Prochaska97}\citeANP{Prochaska98} (1997, 1998),
have argued that the observed high velocity widths 
($\sim$ 300 \kms) and asymmetry of
DLA line-profiles are indicative of a 
population of rapidly rotating thick discs, which
are incompatible with the slowly rotating exponential 
discs modelled in semi-analytic schemes 
(see \citeANP{Kauffmann96} 1996).
However, \citeANP{Haehnelt98} (1998) show
that the same observations can be reproduced
by modelling DLAs as gaseous systems which 
are supported by a mixture of random motions, merging, 
infall $and$ rotation (see also \citeANP{Maller99} 1999; 
\citeANP{Ma01} 2001). Such agglomerations, with virial velocities of
$\sim$ 100 \kms, are consistent 
with the PGD kinematics assumed in \Galform.

In creating the metal-poor GCs, we found it necessary
to truncate blue GC formation in PGDs at high
redshift, in something of an {\it ad-hoc} manner.
Since this occurs when the gas surface-densities of these
PGDs are still high and rapidly forming 'unbound' stars. it begs
the question what physical mechanism could possibly halt GC
formation so abruptly?

One possibility arises from the cosmic re-ionisation
(\eg see \citeANP{VandenBergh01} 2001).
At some point in the early Universe, gas cooled and 
began forming stars, providing enough $UV$ flux
to ionise the neutral inter-galactic medium (e.g 
\citeANP{Loeb01} 2001).
Recent studies of QSO absorption systems have
possibly now constrained the onset of this re-ionisation to 
be 5 $\leq z \leq$ 6 (\citeANP{Fan00} 2000; \nocite{Becker01} 
Becker \etal 2001; \citeANP{Djorgovski01} 2001).
\citeANP{Cen01} (2001) has suggested that such an
external radiation field is sufficient to
compress haloes of mass $\leq$ 10$^{7.5}\Msun$, 
thereby forming self-gravitating baryonic systems,
under the proviso that such haloes have low
initial angular momentum.
By assuming that such systems fragment and
form stars, \citeANP{Cen01} (2001) shows that 
these stellar systems will have
characteristic masses and sizes consistent
with metal-poor GCs. 

Whilst plausible, this explanation is rather
unsatisfactory.
Given our current knowledge that GCs are forming 
in a wide range of star-forming environments, the assertion that 
the formation of metal-poor GCs is somehow a 'special' process 
in the early Universe seems somewhat contrived. 
Perhaps a more self-consistent picture of GC formation
is suggested by the study of \citeANP{Larsen00}
(2000). These authors found that the efficiency of star cluster 
formation is a function of the local SFR in spiral discs,  
implying that the efficiency of GC formation may
generically depend upon the rate of star formation 
in their host galaxy.
In this way, GCs may represent the long-lived tracers  
of the peaks of the cosmic SFR, occurring in relatively 
un-evolved PGDs at high redshift.
Unfortunately, from the standpoint of this study,
we found that an $\epsilon \propto$ SFR relation
fails to reproduce a metal-poor GC component anything 
like that observed. This arises because the SFR in our model 
peaks at $z = 1 \sim 2$, rather than at the required $z > 5$ 
(\ie greater than our blue GC truncation 
redshift).

However, whilst contemporary observations of the cosmic SFR 
indicate that it reaches a peak at $z \leq 2$ 
(\eg see compilation of \citeANP{Blain99}
1999), its behaviour at
higher redshifts ($z > 4$) is relatively unconstrained
\cite{Loeb01}.
\citeANP{Barkana00} (2000) found
from CDM simulations that the comoving SFR 
should exhibit a distinct decrease around the re-ionisation 
redshift (\eg see their figure 1, for the case
of $z_{\rm reion} = 7$). A higher SFR at such redshifts 
could plausibly give rise to the formation 
of significant numbers of blue GCs.

\section{Summary and Conclusions}
\label{sec:SummaryandConclusions}

We have outlined a scheme for forming the 
globular cluster (GC) systems of elliptical galaxies 
within the framework of a semi-analytic model
of galaxy formation. Our principal results are as follows:

\begin{itemize}
\item we can reproduce the observed bimodal colour 
distributions of GC systems associated 
with elliptical galaxies by assuming {\it i)} metal-rich globular
clusters are formed during gaseous mergers and {\it ii)} 
metal-poor GCs are formed in gaseous proto-galactic
discs, and this formation is truncated at $z \sim$ 5.
\item gaseous mergers which lead to the formation of 
metal-rich GCs and field stars do not significantly effect the
total $\SN$ of the host elliptical galaxy. This remains true 
even if a relatively high formation efficiency  is 
assumed ($\epsilon_{\rm red} = 0.007$, assuming no dynamical 
destruction of the GCs occurs.)
\item the number of GCs ($N_{\rm GC}$) scales with 
host galaxy luminosity ($L_{V, \rm gal}$) as $N_{\rm GC} \propto L_{V, \rm
gal}^{1.25}$. This arises due
to an increasing $\mathcal{M}/L$ ratio of the
simulated ellipticals with luminosity. 
This is consistent with observations of 
luminous 'normal' ellipticals, but inconsistent 
with observations of the highest-$\SN$ galaxies.
\item the mean colours of both the metal-rich and metal-poor
GCs weakly correlate with the 
magnitude of the host galaxy, a result is consistent with the 
current observations.
\item the mean colours of the metal-rich GCs are
very similar to the mean colours of their host galaxy. The
colours of the metal-poor GCs are significantly bluer than their
host galaxy, corresponding to a metallicity difference of 
$\sim$ 1.0 dex. This is consistent with contemporary
observations.
\item we find that 93\% of the simulated GC systems are
intrinsically bimodal. However, whether this is 'detected'
in broad-band colours is extremely sensitive to sample size, 
photometric uncertainties, and to a lesser extent, the
statistical test employed to detect sub-populations.
\item the metal-poor GCs have mean ages of
$\sim$ 12 \Gyr, whilst the metal-rich GCs have mean
ages of $\sim$ 9 \Gyr. The age range of the metal-rich GCs 
in the majority of individual galaxies
is large (5 $\sim 12$ \Gyr). This age range 
increases for low-luminosity ellipticals, and 
for ellipticals in lower mass dark 
matter haloes. 
\end{itemize}

We have found that by assuming GC formation
occurs contemporaneously with unbound star formation, 
two phases of GC formation are required to produce
the observed GC systems of elliptical galaxies.
Hierarchical merging at high redshift results
in a metal-rich peak consistent with observations,
if the metal-rich GC formation efficiency 
is $\sim$ 0.7 \% by stellar mass. 
Formation of the metal-poor peak requires
less efficient ($\sim$ 0.2 \%)
GC formation in low-mass haloes at $z \geq 5$.
Models which represent a synthesis of 
star formation, galaxy formation and cosmology
will be required to determine their origin.

The decoupling of the formation of the
metal-poor GCs from the galaxy 
stars at high redshift implies that they
are not direct probes of the galaxy star formation. 
Rather, the metal-poor GCs follow the star formation in relatively 
un-enriched gas associated with low-mass haloes at early times
(\eg \citeANP{Ashman93a} 1993).
In contrast, the metal-rich GCs directly trace the 
merger history of their host galaxy whenever
star formation occurs.
Therefore, the metal-rich GCs are remnants of both 
star formation at later times, and the dynamical 
assembly of their host galaxies through merging. 
We suggest that the metal-rich GCs provide a key 
to probing the formation epoch of elliptical galaxies.

\section{Acknowledgements}

We would like to thank Brad Gibson, Bill Harris,
Gretchen Harris and the anonymous referee 
for extremely useful comments regarding this paper.
We acknowledge KFF99 for the use 
of their stellar population models, and the use of the 
STARLINK facilities at the University of Durham.
MB would like to thank the Royal Society for its
fellowship grant, MONBUSHO
for its excellent research exchange programme and Nobuo Arimoto 
for insightful discussions during 
the early stages of this study at NOAO. 

\bibliographystyle{mnras}
\bibliography{mnras}    
     
\end{document}